\documentclass[a4paper,11pt]{article}
\usepackage{jcappub}
\usepackage[T1]{fontenc}

\usepackage[section]{placeins}
\usepackage{amsmath, amssymb,braket}
\usepackage{color}
\usepackage{float}
\usepackage{graphicx}
\usepackage{subcaption}
\usepackage[numbers,sort&compress]{natbib}
\usepackage{adjustbox}
\usepackage[colorlinks=true,citecolor=blue,urlcolor=blue]{hyperref}
\usepackage{comment}
\usepackage{eso-pic}
\usepackage{tikz}

\newcommand\orcid[1]{\href{http://orcid.org/#1}{\adjustbox{trim={-.15\width} {0\height} {-.15\width} {0\height},clip}{\includegraphics[height=10pt]{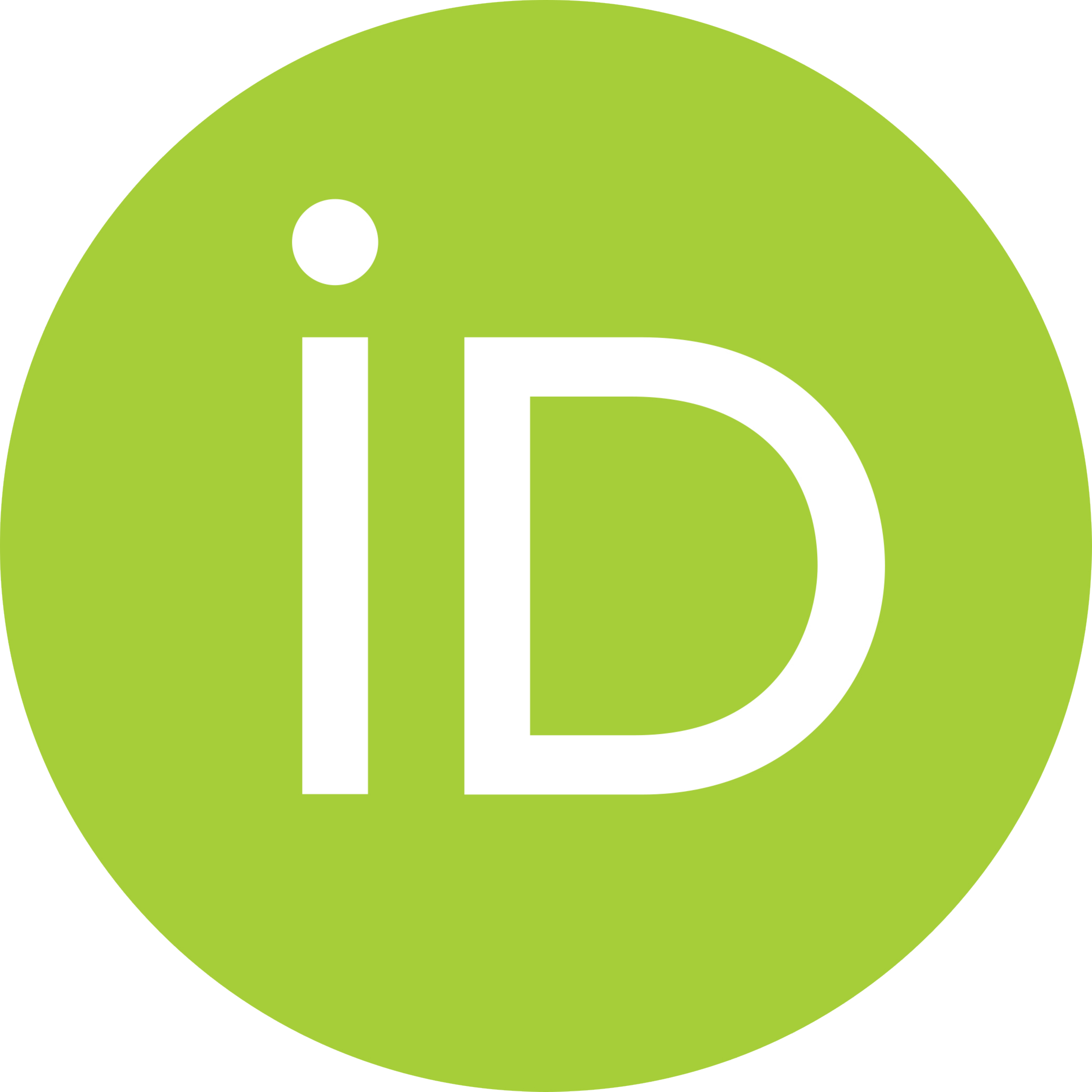}}}}

\begin{document}

\title{
    Primordial or Post-Inflationary: Distinguishing the Origin of Scale-Dependent Features in the CMB
}

\begin{flushright}
\texttt{MIT-CTP/6141},

\texttt{CERN-TH-2026-280}
\end{flushright}

\author[a]{Matteo Braglia\orcid{0000-0003-3421-4069}, }
\author[b,c,d]{Wenzer Qin\orcid{0000-0001-7849-6585}, }
\author[e]{\\ Xingang Chen\orcid{0000-0002-1071-6526}, }
\author[c,d]{Tracy R.~Slatyer\orcid{0000-0001-9699-9047}}

\affiliation[a]{Theoretical Physics Department, CERN, 1211 Geneva 23, Switzerland}
\affiliation[b]{Center for Cosmology and Particle Physics, Department of Physics, New York University, New York, NY 10003, USA}
\affiliation[c]{Department of Physics, Massachusetts Institute of Technology, Cambridge, MA 02139, USA}
\affiliation[d]{Center for Theoretical Physics -- a Leinweber Institute, Massachusetts Institute of Technology, Cambridge, Massachusetts 02139, USA}
\affiliation[e]{Institute for Theory and Computation, Harvard-Smithsonian Center for Astrophysics, 60 Garden Street, Cambridge, MA 02138, USA}

\abstract{
    \noindent Physics beyond the Standard Model, whether of primordial or post-inflationary origin, can leave scale-dependent signals in the anisotropies of the Cosmic Microwave Background (CMB).
    In this work, we study how distinguishable these broad classes of signals are and quantify their level of degeneracy.
    In this initial investigation, for primordial features, we limit ourselves to those of the ``sharp'' and ``resonant'' kinds; for post-inflationary features, we consider two qualitatively distinct models: exotic energy injection, which deposits energy into the primordial plasma and modifies the global thermodynamic history, and early dark energy (EDE), which affects the CMB through the background expansion and gravitational perturbations.
    By studying representative examples, we find that while primordial features can leave oscillatory signatures in the CMB power spectrum with essentially arbitrary frequency and some characteristic running, the post-inflationary models considered here produce qualitatively different patterns that cannot generically reproduce the characteristic running of primordial feature oscillations.
    We verify these findings first with a principal component analysis, and show using only the most significant principal components that there are no significant degeneracies between the primordial and post-inflationary features considered here, and secondly with a full likelihood analysis, which shows that the resulting posteriors do not change significantly when considering individual classes of models or combinations of primordial/post-inflationary features.
    While the two post-inflationary models probe distinct physical mechanisms and both are implemented in a simplified, phenomenological fashion, the methods used in this work can be generalized to include more realistic signals or other classes of signals.
}

\maketitle

\section{Introduction}
\label{sec:intro}
The Cosmic Microwave Background (CMB) is a cornerstone of modern cosmology and has set some of the most stringent constraints on the $\Lambda$CDM cosmological parameters. 
The detection of CMB polarization has further broken degeneracies between primordial physics and the later epoch of reionization. 
These parameters are tightly constrained through observations of the CMB anisotropies by \textit{Planck}~\cite{Planck:2018vyg}, and more recently by high-resolution ground-based experiments such as ACT~\cite{ACT:2025fju,ACT:2025tim} and SPT~\cite{SPT-3G:2025bzu}, as well as complementary large-scale structure (LSS) observations from e.g. BOSS~\cite{BOSS:2012dmf} and DESI~\cite{DESI:2025zgx}\footnote{Some tensions remain between CMB and other datasets, notably the $H_0$ and $S_8$ tensions, although recent data analyses from KiDS have considerably softened the latter tension~\cite{Wright:2025xka,Pantos:2026koc}. In addition, recent BAO measurements from DESI, combined with supernova datasets, could challenge the standard model's description of late-time cosmic evolution.
Despite these questions, $\Lambda$CDM remains the best description of our Universe, with no definitive evidence yet favoring alternatives.}.

Yet, measurements of the CMB anisotropy power spectra reveal residuals that may not be fully explained within the $\Lambda$CDM framework. Some of these appear across multiple datasets, the most notable being the so-called low-$\ell$ anomaly: a deficit of power in the large-scale temperature anisotropies observed independently by both WMAP and Planck \cite{Peiris:2003ff,Akrami:2018odb}. There are also some noticeable glitches at high-$\ell$ that have been analyzed in the literature ~\cite{Akrami:2018odb,Planck:2018vyg}. While such residuals are not statistically significant and may simply reflect statistical fluctuations or unidentified systematics, they could also be early hints of physics beyond the standard model of cosmology. Future experiments with higher sensitivity will be crucial in confirming or ruling out this possibility.

In particular, these residuals may be related to \textit{primordial features}, i.e.\ signals predicted by many realistic models of inflation that arise when the assumptions of single-field slow-roll---the vanilla model of inflation---are violated. Such features manifest as strongly scale-dependent, often oscillatory, modifications to the primordial power spectrum of scalar fluctuations. These imprints are transferred to the angular power spectrum of the CMB anisotropies, and to other cosmological observables, thereby offering a potential link between CMB residuals and new energy scales, heavy fields coupled to the inflaton, or even model-independent evidence for or against cosmic inflation itself. See \cite{Chen:2010xka,Chluba:2015bqa,Slosar:2019gvt, Achucarro:2022qrl} for reviews.

Several primordial-feature candidates have been proposed as explanations for the observed residuals, in some cases providing improved fits to the CMB data at both large and small scales \cite{Peiris:2003ff,Akrami:2018odb, Adams:2001vc, Bean:2008na, Mortonson:2009qv, Hazra:2010ve, Hazra:2014goa, Miranda:2014fwa, Hazra:2014jwa, Canas-Herrera:2020mme, Braglia:2021ckn, Braglia:2021sun, Braglia:2021rej, Braglia:2022ftm}. However, no definitive evidence has been established so far, as these models typically require additional parameters beyond $\Lambda$CDM and are therefore penalized in Bayesian model comparison. 
Future CMB experiments will be able to test these scenarios with much higher statistical significance, potentially confirming or ruling out specific feature models. Moreover, they may allow us to discriminate between different primordial mechanisms that could generate the observed signals \cite{Braglia:2022ftm,Petretti:2024mjy}.\footnote{While this paper focuses on CMB data, the case for primordial features is broader, as they also leave correlated imprints in LSS. Current LSS constraints are competitive with CMB limits only for a limited range of feature frequencies, but are expected to surpass them in the near future~\cite{Huang:2012mr,Hazra:2012vs,Chen:2016vvw,Ballardini:2016hpi,Palma:2017wxu,LHuillier:2017lgm,Ballardini:2017qwq,Beutler:2019ojk,Ballardini:2019tuc,Debono:2020emh,Li:2021jvz,Ballardini:2022wzu,Chandra:2022utq,Chen:2026gxp}.}

On the other hand, non-standard physics operating after the inflationary era could also be responsible for the observed CMB residuals. Examples include energy injection from decaying or annihilating dark matter~\cite{Adams:1998nr, Chen:2003gz, Padmanabhan:2005es, Slatyer:2009yq, Kanzaki:2009hf, Slatyer:2015jla, Slatyer:2016qyl, Poulin:2016anj, Cang:2020exa, Capozzi:2023xie}, evaporation and accretion onto primordial black holes~\cite{Stocker:2018avm, 2022JCAP...03..030M}, as well as interactions between dark matter and dark radiation, additional energy components such as early dark energy (EDE)~\cite{Karwal:2016vyq,Poulin:2018cxd,Lin:2019qug,Agrawal:2019lmo,Smith:2019ihp,Niedermann:2019olb,Poulin:2023lkg,Lin:2020jcb}, modifications to gravity that alter the expansion history at early times~\cite{Lin:2018nxe,Rossi:2019lgt,Ballesteros:2020sik,Braglia:2020iik,Braglia:2020auw,Adi:2020qqf,FrancoAbellan:2023gec}, or changes in the physics of recombination~\cite{Lee:2022gzh,ACT:2025tim}. 
While primordial features affect the {\em initial conditions}, these post-inflationary scenarios modify the {\em transfer functions}, usually through sub-horizon processes.  
To properly assess cosmological constraints on primordial features, and to forecast their detectability with future experiments, it is therefore crucial to determine under which conditions a primordial feature signal could be misinterpreted as new post-inflationary physics, and vice versa.
The existing literature contains very few systematic investigations of this degeneracy.

Motivated by this, in this paper we take a first step towards addressing this question by analyzing models with exotic energy injections at different cosmological epochs and comparing them to the effects of primordial features. 
We begin in Section~\ref{sec:features} by reviewing the physics of primordial features in Section~\ref{sec:primordial}, focusing in particular on the ``sharp'' and ``resonant'' models which can give rise to oscillatory features of arbitrary frequency in the CMB power spectrum, and then show in Section~\ref{sec:late} that post-inflationary features in the form of exotic energy injection and transient EDE components can only impact the CMB anistropies mainly by effectively increasing the optical depth or shifting the acoustic peaks, which leads to a phase shift in the power spectrum.
In Section~\ref{sec:examples}, we demonstrate this qualitative picture with studies of some fiducial models.
In Section~\ref{sec:PCA} we verify these results using a Principal Component Analysis (PCA), and show that the dominant directions in parameter space associated with primordial and post-inflationary features are nearly orthogonal. 
The only exception is a small subset of primordial-feature models whose characteristic frequencies overlap with those of BAO, leading to mild correlations with exotic energy injections around recombination. 
Even in this case, because there is a non-trivial transfer function linking primordial features to CMB features, the primordial oscillation signal is deformed and its degeneracy with post-inflation features of same frequency is also partly broken.
Finally, in Section~\ref{sec:MCMC}, we perform an analysis of the Planck CMB anisotropy data, finding results consistent with our PCA conclusions.
We conclude in Section~\ref{sec:conclusion}.

\section{Scale-dependent features}
\label{sec:features}

In this Section, we introduce and motivate our models of primordial and post-inflationary features.
We work with simplified examples, which, as such, do not cover the full range of possible features in the CMB residuals, and rather serve as illustrative examples that capture the main characteristics of each category of models.
Moreover, for those interested in investigating the distinguishability between primordial features and other classes of post-inflationary features, the methods described in this work can be straightforwardly generalized to include or substitute for these other features given e.g. a modified version of \texttt{CLASS} or another Boltzmann code that can calculate the impact on the CMB power spectrum.
We leave such generalizations as a direction for future study.

\subsection{Primordial features}
\label{sec:primordial}

Primordial features are strongly scale-dependent corrections to the otherwise nearly scale-invariant scalar power spectrum, produced in the primordial universe by mechanisms such as inflation. They arise from small deviations of the inflationary background from slow-roll (SR) dynamics. Such deviations can occur when the inflationary potential, the couplings between the inflaton and other fields, or the internal field space contain small corrections---such as ripples, bumps, or steps---that transiently or periodically violate the SR assumptions. While the Universe continues to inflate, SR parameters can become transiently or periodically much larger than their slow-roll values.
Depending on the nature of the SR violation, one can broadly distinguish between two categories of features, each leaving distinct imprints in the primordial correlation functions and, consequently, in cosmological observables \cite{Chen:2010xka,Chluba:2015bqa,Slosar:2019gvt,Achucarro:2022qrl}. 

The first category is that of {\em sharp features}. These correspond to transient---shorter than one $e$-fold---departures from SR. These features generate oscillatory corrections to the primordial power spectrum, which can be schematically represented, in term of the fractional correction to the leading order power spectrum $\mathcal{P}_0$, by the following simplified template proposed to only capture the leading running behavior \cite{Chen:2008wn}\footnote{The factor of $2$ in Eq.~\eqref{eq:sharp} applies specifically to the two-point function. For an $n$-point function one would replace $2k$ with $\sum_{i=1}^nk_i$ in the schematic form.}  
\begin{equation}
\label{eq:sharp}
{\rm {\bf Sharp\, features:}}\quad\quad\quad\frac{\Delta \mathcal{P}_{\rm sharp}(k)}{\mathcal{P}_0}= A_{\rm lin} \sin \!\left(\frac{2k}{k_0}+\phi_{\rm lin} \right),
\end{equation}
where $k$ represents wavenumber, $k_0$ is the mode crossing the horizon at the time of the feature, $\phi_{\rm lin}$ is a constant phase, and $A_{\rm lin}$ is a constant amplitude.

The sharp-feature mechanism was introduced by Starobinsky by studying the effect of a sudden change in the slope of a linear potential in single-field canonical inflation \cite{Starobinsky:1992ts}. Subsequent work explored various other types of feature forms, including steps or bumps in the potential or kinetic terms in single-field inflation e.g.~\cite{Adams:2001vc, Chen:2006xjb, Bean:2008na, Mortonson:2009qv, Hazra:2010ve, Hazra:2014goa, Miranda:2014fwa, Hazra:2014jwa, Canas-Herrera:2020mme}, or sudden turns in the inflaton rolling trajectory in multi-field inflationary scenarios e.g.~\cite{Achucarro:2010da,Gao:2012uq}. Note that the simplified template~\eqref{eq:sharp} extends to infinitely large $k$, but this is only valid in the limit of an infinitely sharp, i.e.~instantaneous, feature. For a feature of finite duration, the resulting signal becomes localized in wavenumber space: the amplitude $A_{\rm lin}$ becomes an {\em envelope} $A (k)$ that only has finite support in the $k$-space. In either case, both the envelope and the phase $\phi$ can have mild $k$-dependence, which depends on the underlying realization of the sharp feature and are highly model-dependent.

Despite this model dependence, the {\em running}---the leading-order oscillatory behavior which is linear in $k$---is a universal property of sharp feature signals. A detection of such an oscillatory pattern and its frequency $k_0$ would provide evidence for a primordial sharp feature, as well as information about the timing when the SR violation occurred. For this reason, in this work we focus on this robust property and adopt Eq.~\eqref{eq:sharp} as a toy model, neglecting other milder scale-dependence. We caution, however, that infinitely sharp features are unphysical: an instantaneous discontinuity would inject an infinite energy scale into the inflationary background, pushing the EFT description beyond validity and requiring the inclusion of new physics (see e.g.~\cite{Bartolo:2013exa,Cannone:2014qna,Adshead:2014sga,CarrilloGonzalez:2025fqq,Braglia:2026fle}).

The second category is that of {\em resonant features}. Unlike sharp features, the source for resonant features is not localized in time. Instead, they arise when some background quantity during inflation acquires a small periodic correction of the form $\propto \sin(\omega t)$, where $t$ is cosmic time and $\omega$ represents frequency. Each mode of the inflaton quantum fluctuations temporarily resonates with these oscillations around the time when $k = a \omega$. As long as $\omega > H$, with $H$ being the Hubble rate during inflation, this resonance occurs before horizon crossing $k = a H$. Since the condition $k=a\omega$ is satisfied at different times for different modes, the phase of the background oscillations is repeated multiple times.   As a result, the correction to the power spectrum is captured by the following template~\cite{Chen:2008wn,Flauger:2009ab}
\begin{equation}
\label{eq:resonant}
{\rm {\bf Resonant\, features:}}\quad\quad\quad\frac{\Delta \mathcal{P}_{\rm res}(k)}{\mathcal{P}_0}= A_{\rm log} \sin \!\left(\frac{\omega}{H} \ln k + \phi_{\rm log} \right) ~,
\end{equation}
where $A_{\rm log}$ is the amplitude of this feature, determined by the resonance mechanism, and $\phi_{\rm log}$ is a constant phase.
In this case, the {\em running} is logarithmic in $\ln k$-space, in contrast to sharp features where the running is linear in $k$-space, producing distinct observational signatures.  

The resonance mechanism was introduced phenomenologically in Ref.~\cite{Chen:2008wn}, and it was soon realized by more UV-completed models, e.g.~axion monodromy, in which a softly broken discrete shift symmetry of an axion induces oscillations of constant amplitudes in the inflaton potential \cite{Flauger:2009ab,Flauger:2010ja}.
In this class of models, the correction to the power spectrum at the leading order takes the form of Eq.~\eqref{eq:resonant}, with a constant amplitude $A_{\rm log}$.  
As with sharp features, the most informative parameter to extract observationally is the frequency $\omega/H$, which directly encodes the presence of a new physical scale during inflation.

In addition, we note that there are feature models whose signals are nontrivial combinations of sharp and resonant features. An important example is the so-called {\em classical primordial standard clock} models \cite{Chen:2011zf,Chen:2014cwa,Quintin:2024boj}\footnote{Classical Primordial Standard Clocks are related, but different from Quantum ones \cite{Chen:2015lza,Chen:2018cgg}. The latter, which are the target of the so-called Cosmological Collider Physics program~\cite{Chen:2009zp,Baumann:2011nk,Noumi:2012vr,Arkani-Hamed:2015bza}, are massive fields that oscillate only at the level of the perturbations, but {\em not} at the level of the background level, therefore leaving the resonance imprints only on the shape-dependence of $n>2$ correlation functions.}. In this scenario, a sharp feature excites the classical oscillation of a massive field with mass $m \gg H$, generating a hybrid of the two kinds of signals. In particular, the resonant contribution in the inflation scenario acquires a decaying envelope, reflecting the relaxation of the scalar field oscillations around the (effective) minimum of its potential. The frequency $\omega$ is set by the mass $m$ of the massive field, offering a unique possibility to directly probe massive fields active during inflation, and thereby obtain evidence for the multifield nature of the inflationary dynamics.  
These models also provide a way to distinguish inflation from alternative primordial-Universe scenarios, such as contracting or ekpyrotic models. Indeed, Eq.~\eqref{eq:resonant} implicitly assumes an inflationary background of the form $a(t)\propto t^p$ with $p\to\infty$, reproducing exponential expansion. In more general power-law scenarios, the clock signal instead takes the form $\Delta \mathcal{P}_{\rm clock}/\mathcal{P}_0 \sim
\sin [p\,(m/H) (2k)^{1/p}]$~\cite{Chen:2011zf}. A determination of the index $p$ would therefore offer a direct test of inflationary expansion, complementary to the detection of primordial tensor modes.

Finally, besides mechanisms introduced by features in inflation models, oscillatory signals in the primordial spectra may also be induced by tran-Planckian physics or non-standard initial conditions (e.g.~\cite{Brandenberger:2000wr,Easther:2002xe,Burgess:2002ub,Martin:2003sg}).

To summarize, measuring a primordial feature signal and constraining the parameters that characterize it would provide a unique observational window into microphysical physics beyond the simplest slow-roll paradigm. The sensitivity of current and future data allows nontrivial constraints on these parameters and, to some extent, the ability to identify or test best-fit candidates and compare different feature models \cite{Braglia:2022ftm}. 

Although it is important to model primordial features accurately, given the current lack of evidence for features, as well as the exploratory nature of our study---in particular, the fact that prospects of the degeneracy of primordial and post-inflationary features has not been systematically investigated before---we adopt a simplified approach. Specifically, we consider the two templates in Eq.~\eqref{eq:sharp} and Eq.~\eqref{eq:resonant} as representative cases, and we do not include more complicated models. In fact, adding further details to the primordial feature models would in general make the corresponding signals more distinguishable, so our choice should be viewed as overall conservative.

\subsection{Post-inflationary features}
\label{sec:late}

{\bf Exotic energy injection.} We use exotic energy injection as the first example of post-inflationary features that can imprint on the CMB.
Exotic energy injection refers to a transfer of energy into (typically electromagnetic) observables by exotic processes and is a prediction of a broad range of theories beyond the Standard Model.
For example, energy injection from dark matter annihilation and decay can impact the CMB power spectrum~\cite{Adams:1998nr, Chen:2003gz, Padmanabhan:2005es, Slatyer:2009yq, Kanzaki:2009hf, Slatyer:2015jla, Slatyer:2016qyl, Poulin:2016anj, Cang:2020exa, Capozzi:2023xie}, CMB blackbody spectrum~\cite{Chluba:2013vsa,Ali-Haimoud:2015pwa,Chluba:2015hma,Acharya:2018iwh,Liu:2023fgu,Liu:2023nct}, IGM temperature~\cite{Cirelli:2009bb, Diamanti:2013bia, Liu:2016cnk, Liu:2020wqz, Capozzi:2023xie}, and 21\,cm cosmology signals~\cite{Evoli:2014pva, Lopez-Honorez:2016sur, DAmico:2018sxd, Liu:2018uzy, Cheung:2018vww, Mitridate:2018iag, Clark:2018ghm}, as can evaporation by or accretion onto primordial black holes~\cite{Stocker:2018avm, 2022JCAP...03..030M}.

We focus on homogeneous exotic energy injection---while the impact of inhomogeneities on the energy deposition cascade are important for later observables e.g. within 21\,cm cosmology, they are unlikely to have a strong impact on the CMB signals~\cite{Dvorkin:2013cga,Liu:2019bbm,Sun:2023acy}. 
Under the assumption of homogeneity, one can parametrize the effects through energy deposition functions $f_c (z)$, which are defined by
\begin{equation}
    \left( \frac{dE}{dV \, dt} \right)^\mathrm{dep}_c = f_c(z) \left( \frac{dE}{dV \, dt} \right)^\mathrm{inj} ,
\end{equation}
where $c$ denotes a deposition channel such as heating, hydrogen ionization, etc. 
The $f_c (z)$ functions can be calculated using codes such as \texttt{DarkHistory}~\cite{Liu:2019bbm,Liu:2023fgu,Liu:2023nct} or its precursors~\cite{Galli:2013dna}.
In principle, the $f_c (z)$ functions depend on the energies and species of the injected particles; however, to study the impact of energy injection in a relatively model-independent manner and avoid specifying the outgoing particles, we will use the simple prescription of Refs.~\cite{Shull:1982zz,Chen:2003gz}, given by 
\begin{equation}
    f_\mathrm{ion} (z) = f_\mathrm{exc} (z) = \frac{1 - x_e (z)}{3}, \qquad f_\mathrm{heat} (z) = \frac{1 + 2 x_e (z)}{3} ,
    \label{eq:fs}
\end{equation}
where $x_e(z) \equiv n_e(z)/n_{\mathrm H}(z)$ is the free-electron fraction,
with $n_{\mathrm H}$ denoting the total number density of hydrogen nuclei.
The global rate of energy injection is given for annihilating and decaying dark matter by
\begin{align}
    \left( \frac{dE}{dV dt} \right)^\mathrm{inj}_\mathrm{ann} &= p_\mathrm{ann} (z) c^2 \Omega_\mathrm{DM}^2 \rho_c^2 (1+z)^6 , \\
    \left( \frac{dE}{dV dt} \right)^\mathrm{inj}_\mathrm{dec} &= p_\mathrm{dec} (z) c^2 \Omega_\mathrm{DM} \rho_c (1+z)^3 ,
\end{align}
where the $p(z)$ functions are given by $p_\mathrm{ann} (z) = \langle \sigma v \rangle / m_\mathrm{DM}$ and $p_\mathrm{dec} (z) = \Gamma$, with $\langle \sigma v \rangle$ being the thermally averaged cross-section for dark matter self-annihilation, $m_\mathrm{DM}$ being the dark matter mass, and $\Gamma$ being the decay lifetime, $\Omega_\mathrm{DM}$ is the density parameter of dark matter, and $\rho_c$ is the critical density today.

In this work, we will study energy deposition at specific redshifts by setting $\left( \frac{dE}{dV \, dt} \right)^\mathrm{dep} \propto \delta(z - z_\mathrm{inj})$, where $\delta (z)$ denotes the Dirac delta function or its approximation by a narrow Gaussian function.
The rate of energy deposition for each channel is then used to determine the evolution of thermodynamic quantities such as the global matter temperature and free electron fraction, which consequently affect the multipole moments of the CMB power spectrum, $C_\ell$.

With regards to the impact on the CMB power spectrum, the most important effect comes from additional contributions to the global ionization history.
Consequently, while in principle our treatment depends on the on-the-spot approximation in Eq.~\eqref{eq:fs}, because we fix the redshift dependence of the deposition, in practice the effect of using a more complete treatment would just be to rescale the total deposited power going into ionization at that redshift (shifting the overall normalization of the effect on the CMB). Thus we expect our results to hold more generally.
For energy injections near the redshift of recombination, the additional ionization slightly delays recombination, and thus the location of the peaks in the $C_\ell$'s.
For later injections, it is already known that additional ionizations damp the amplitude of the CMB power spectrum by effectively raising the optical depth, $\tau = \int n_e \sigma_T \,dt$, where $n_e$ is the free electron density and $\sigma_T$ is the Thomson cross-section~\cite{Padmanabhan:2005es}.
We discuss these effects in greater detail in Appendix~\ref{app:analytic}.
Hence, the main impact of exotic energy injection on the $C_\ell$'s is either an \textit{overall and mildly $\ell$-dependent damping} due in part to the change in optical depth or a \textit{phase shift} from modified timing of recombination, which appears in the residuals as oscillations with frequency equal to the BAOs.
These effects are also discussed in Ref.~\cite{Padmanabhan:2005es}.
By comparison, the primordial features described in Section~\ref{sec:primordial} can appear in the CMB power spectrum as residuals with, in principal, any frequency.

{\bf Transient EDE.} As a different example of post-inflationary feature model, we also consider a localized modification to the background dynamics, in the form of a brief increase in the expansion rate of the Universe, i.e. $\Delta H(z)>0$. In addition to providing a qualitatively different mechanism for producing scale-dependent CMB residuals, this class of models is motivated as they are a promising mechanism to resolve the so-called Hubble tension (see e.g.~\cite{DiValentino:2020zio} for a review) when the increase in the expansion rate occurs around the time of recombination. This mechanism is easily implemented within early dark energy (EDE) models~\cite{Karwal:2016vyq,Poulin:2023lkg}, where a scalar field is frozen by the Hubble friction at early times effectively behaving as dark energy. When its (effective) mass becomes comparable to the Hubble radius, the scalar field starts rolling down the potential and eventually undergoes damped oscillations around the minimum of the potential, whose structure is chosen in such a way that the scalar field energy density redshifts away faster than radiation. In this way, the scalar field only contributes to the dynamics of the Universe for a limited amount of time, centered around the redshift at which its velocity is maximum, and shortly after it has unfrozen. 

In this work, rather than starting from a specific scalar-field potential, we aim to capture the main phenomenological features of this class of models by working directly at the level of the equation of state (EoS). We model the additional component through a time-dependent EoS that undergoes a transition at a critical redshift $z_\mathrm{EDE}$, from
$w_{\rm EDE}(z\gg z_\mathrm{EDE})=-1$, characteristic of vacuum energy and therefore of an early-dark-energy-like component, to $w_{\rm EDE}(z\ll z_\mathrm{EDE})=w_f$. We parameterize this transition through
\begin{equation}
 w_{\rm EDE}(a)
 = -1 + \frac{1+w_f}{2}
 \left[1-\tanh\!\left(\frac{\ln a_\mathrm{EDE}-\ln a}{\sigma}\right)\right],
 \label{eq:wfld}
\end{equation}
where $a_\mathrm{EDE}=(1+z_\mathrm{EDE})^{-1}$ and $\sigma$ controls the sharpness of the transition in $\ln a$.
This model is closely related to the so-called acoustic dark energy (ADE) scenarios introduced in Refs.~\cite{Lin:2019qug,Lin:2020jcb}, and shares the same qualitative behavior. Here, however, we adopt the different EoS parameterization in Eq.~\eqref{eq:wfld}, which makes the transition between the two asymptotic regimes particularly transparent.

Given the functional form in Eq.~\eqref{eq:wfld}, the evolution of the EDE energy density follows from the continuity equation,
\begin{equation}
 \dot{\rho}_{\rm EDE}+3H(1+w_{\rm EDE})\rho_{\rm EDE}=0,
\end{equation}
and is given by
\begin{equation}
 \rho_{\rm EDE}(a)=\rho_{{\rm EDE},0}\,e^{I(a)},
\end{equation}
with
\begin{align}
 I(a) = \frac{3}{2}(1+w_f)\Bigg[&-\ln a
 -\sigma\Bigg(
 \ln\cosh\!\frac{\ln a-\ln a_\mathrm{EDE}}{\sigma}
 -\ln\cosh\!\frac{-\ln a_\mathrm{EDE}}{\sigma}
 \Bigg)\Bigg],
 \label{eq:Ia}
\end{align}
where $\rho_{{\rm EDE},0}$ denotes the present-day value of the EDE energy density.

At the background level, the model therefore depends {\it a priori} on four parameters, namely $w_f$, $a_\mathrm{EDE}$, $\sigma$, and $\rho_{{\rm EDE},0}$. It is convenient to trade the last parameter for an amplitude $f_{\rm EDE}$ characterizing the fractional contribution of EDE to the critical density at the time of transition. In addition, the model contains the sound speed $c_s^2$, which controls the evolution of its linear perturbations.

In the main text, we restrict ourselves to a simplified realization of this scenario. 
We consider a very sharp transition, fixing $\sigma$ to a small value.
To confine the EDE contribution to a very limited redshift range, we also take a relatively large value for the late-time equation of state, $w_f=7$, so that after the transition the fluid rapidly dilutes away as $\rho_{\rm EDE}\propto a^{-12}$. 
We furthermore assume a canonical sound speed, $c_s^2=1$. Within this simplified setup, the only free parameters are therefore the transition redshift $z_\mathrm{EDE}$
and the amplitude $f_{\rm EDE}$. 
We have tested that a more general analysis in which the remaining parameters are varied does not change our conclusions.

\subsection{Examples}
\label{sec:examples}

\begin{figure*}
    \centering
    \includegraphics[width=0.92\linewidth]{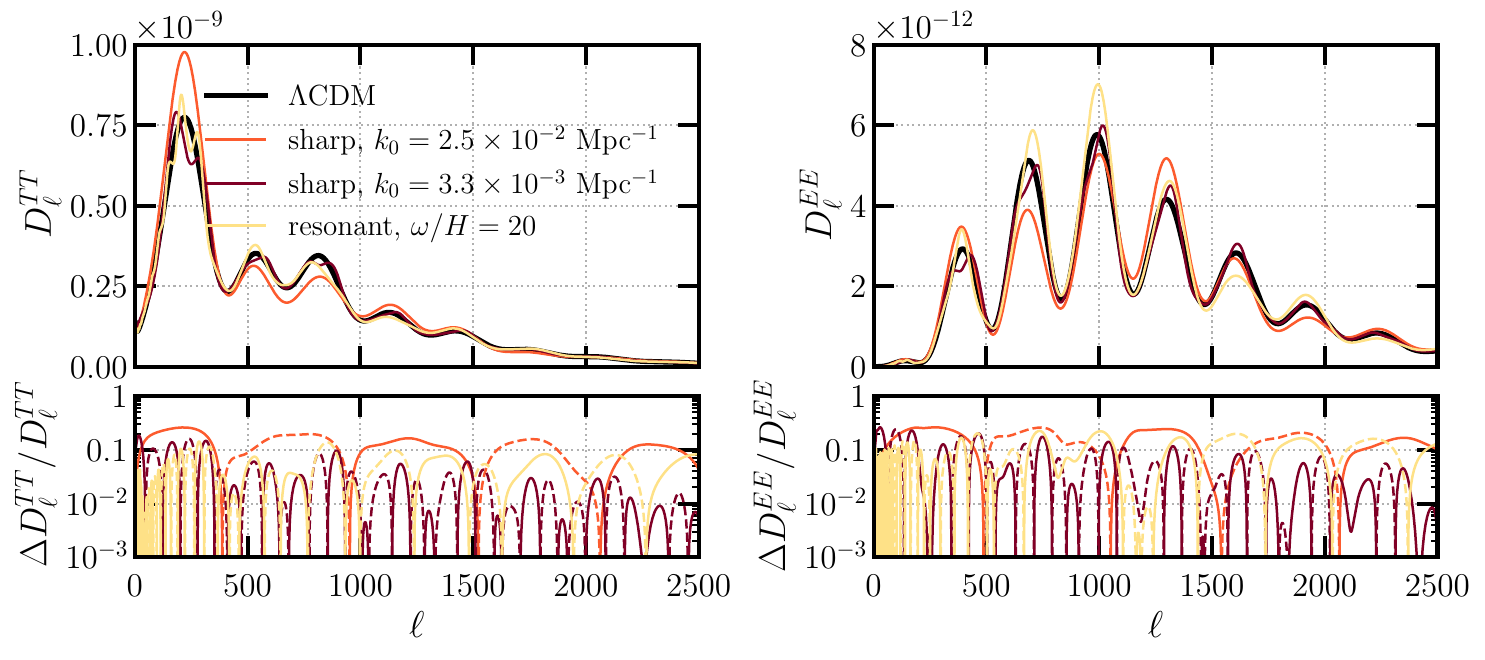}
    \includegraphics[width=0.92\linewidth]{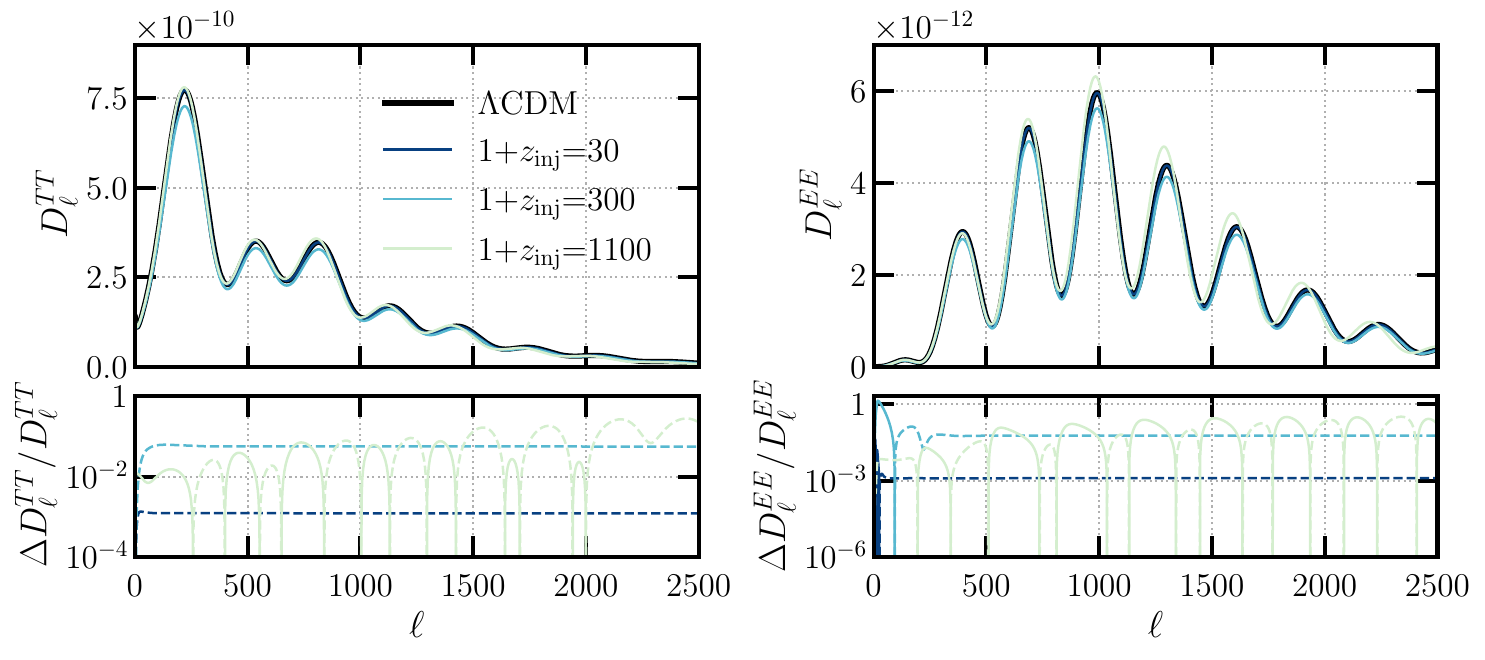}
    \includegraphics[width=0.92\linewidth]{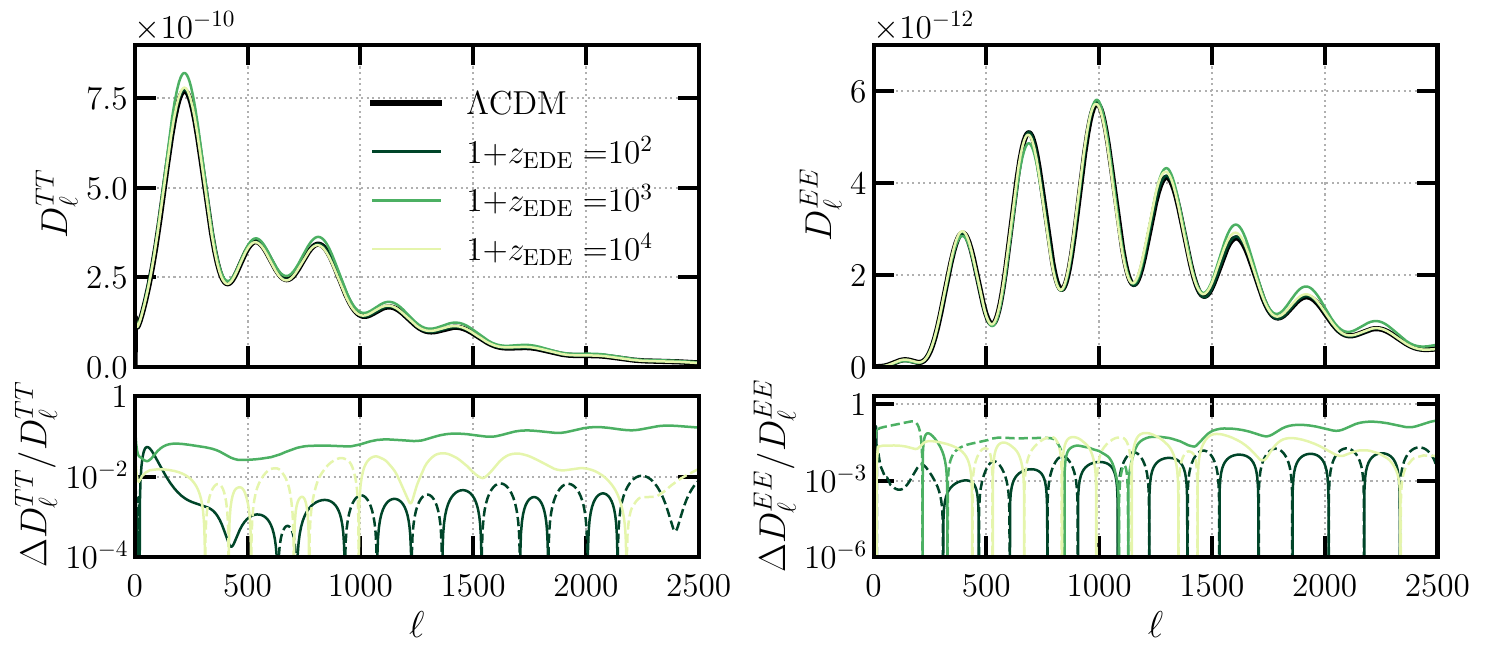}
    \caption{
        The CMB TT (left) and EE (right) power spectra and residuals with respect to $\Lambda$CDM (right) when including different types of features. 
        The top row demonstrates the effect of primordial features, the center shows exotic energy injection, and the last row shows and EDE-like fluid.
        Dashed lines in the lower panels indicate negative values of the residuals. 
    }
    \label{fig:example_spectra}
\end{figure*}
\begin{figure*}
    \centering
    \includegraphics[width=0.92\linewidth]{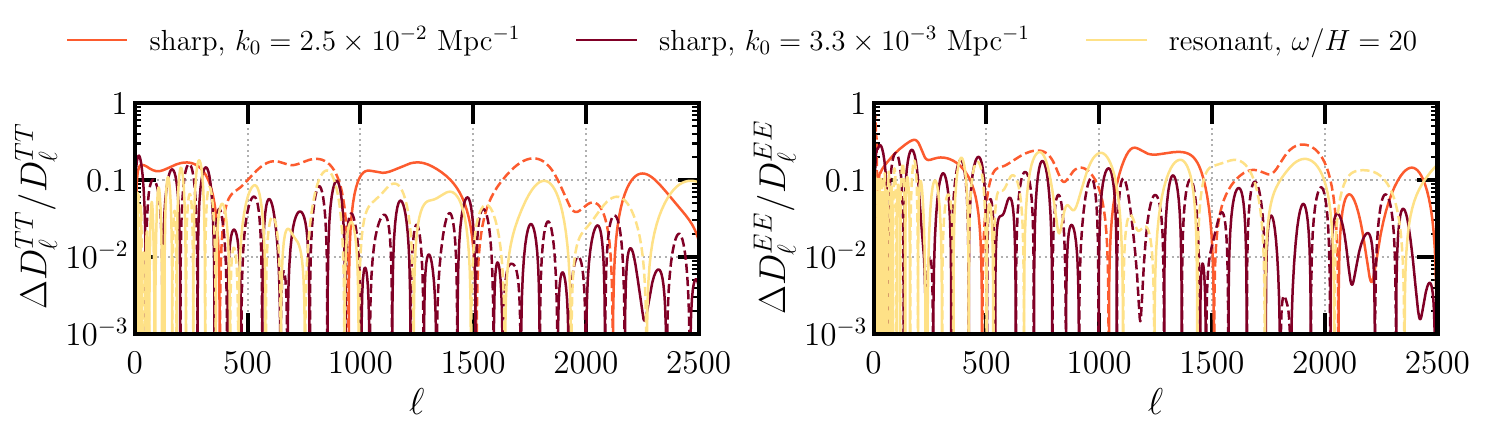}
    \includegraphics[width=0.92\linewidth]{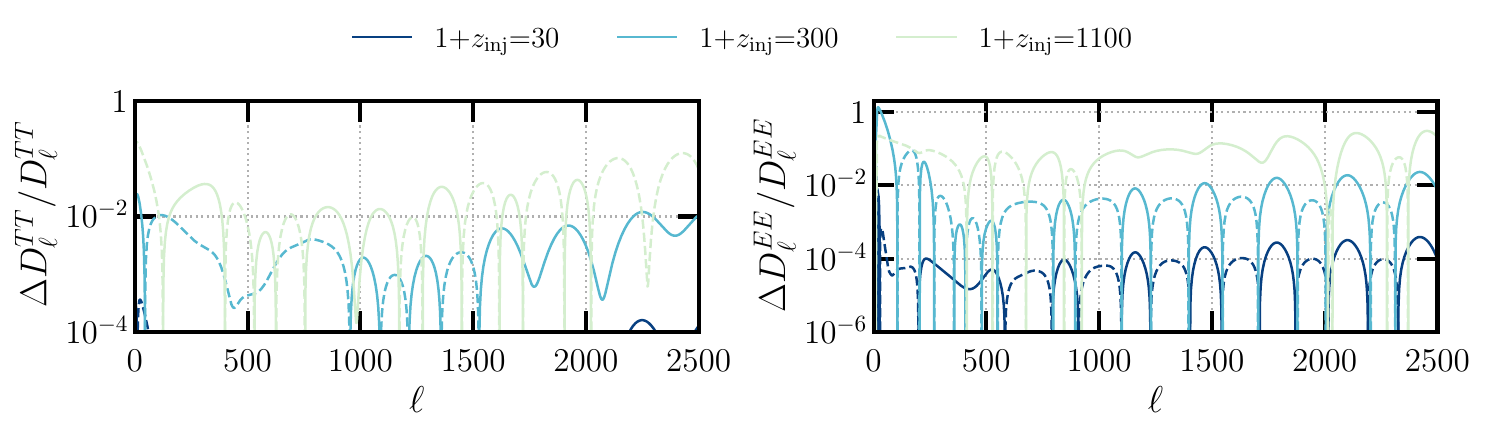}
    \includegraphics[width=0.92\linewidth]{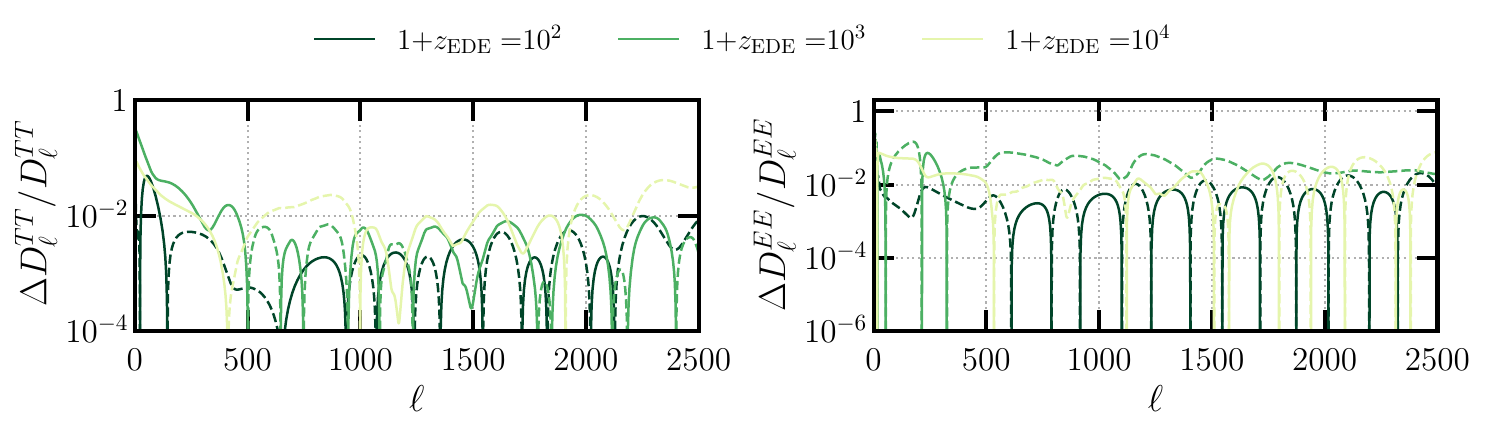}
    \caption{
        The same models as in Fig.~\ref{fig:example_spectra}, but showing the resulting residuals \textit{after} refitting the $\Lambda$CDM parameters.
        Given the large amplitude of the exotic features considered here, the corresponding shifts in the cosmological parameters are similarly large and are considered only for illustrative purposes.
    }
    \label{fig:example_spectra_marg}
\end{figure*}

Having worked out some intuitive results for the various classes of scale-dependent features, we verify these findings by studying some examples with numerical tools.
To this end, we will use a modified version of \texttt{ExoCLASS}, which is a branch of the \texttt{CLASS} Boltzmann code that is designed to include the effects of exotic energy injection~\cite{Stocker:2018avm}.
The primordial features discussed in Section~\ref{sec:primordial} are included through a simple modification of the \texttt{primordial} module that adds features of the form Eq.~\eqref{eq:sharp} and \eqref{eq:resonant}.
To include exotic energy injection, we manually specify the energy deposition fractions within \texttt{ExoCLASS} using the option to include an \texttt{energy deposition function file}---see Appendix B of Ref.~\cite{Stocker:2018avm} for details on this usage.
As mentioned previously, the rate of energy deposition for each channel is then used within \texttt{ExoCLASS} to calculate the rate equations for thermodynamic quantities, which impact the CMB anisotropies. To include the EDE fluid in \texttt{CLASS}, one just needs to implement the functional form of Eq.~\eqref{eq:Ia} in the \texttt{background} module. The changes are automatically propagated to the \texttt{perturbations} module.

Fig.~\ref{fig:example_spectra} shows a few examples of the CMB power spectra resulting from each  class of features.
The top row shows the primordial features described in Section~\ref{sec:primordial}, including sharp features with oscillation frequencies of $k_0 = 2.5 \times 10^{-2}, 3.3 \times 10^{-3}$ Mpc$^{-1}$ and resonant features with frequencies of $\omega/H = 4$ .
The left panel shows the CMB temperature power spectrum for each model, while the right panel shows the polarization (EE) spectra; the lower panels show the differences with respect to $\Lambda$CDM.
One can see that the residuals with respect to the CMB generally reflect oscillations with the corresponding frequency in $\ell$.

The second row of Fig.~\ref{fig:example_spectra} similarly demonstrates the effect of post-inflationary features through energy injections at redshifts of $1+z = 30$, 300 and 1100.
For the injection at $1+z = 1100$, which is at the time of recombination, the main effect is to delay recombination, thus shifting the location of the peaks and therefore the residuals correspond to a phase shift, e.g. oscillations with frequency corresponding to the BAO frequency.
For later injections, the ionizations caused by the exotic injection do not impact the time of recombination, but effectively raise the optical depth, thus causing an overall damping of the power spectrum, which results in flat residuals in the right panel.

As for EDE, for transitions before recombination, the additional energy density increases the expansion rate and reduces the sound horizon and diffusion scale, shifting the acoustic
peaks and producing the characteristic oscillatory pattern in the residuals.
In addition to this dominant background effect, EDE perturbations modify the gravitational potentials and therefore the acoustic driving of the photon-baryon fluid, predominantly for modes entering the horizon around $z_{\rm EDE}$.
This leads to additional localized modifications of the amplitudes of the acoustic peaks, whose characteristic scale depends on the transition redshift~\cite{Poulin:2023lkg}.
For transitions after recombination, such as our lowest-redshift example, the sound horizon is instead essentially unaffected, and the impact arises mainly through changes
in the angular-diameter distance and the evolution of the gravitational potentials. Despite these additional effects, EDE does not introduce a new oscillatory frequency, and the dominant oscillatory structure in the CMB residuals is due to the shift of the acoustic peaks.

While the residuals of Fig.~\ref{fig:example_spectra} conveys how the addition of exotic features alters the CMB power spectrum, they do not show the residual that would be recovered from an actual analysis of the CMB anisotropies, where cosmological parameters that have some degeneracy with the exotic features can absorb some of the signal.
Fig.~\ref{fig:example_spectra_marg} shows the resulting residuals after refitting the $\Lambda$CDM parameters---since the example exotic features have amplitudes that are exaggerated for visibility, the shifts in the cosmological parameters are very large and hence these residuals are only for illustrative purposes.
Since the primordial feature residuals appear with arbitrary frequencies that cannot be easily reproduced by other physics, they remain largely unchanged.
The exotic injection features at $1+z < 1100$, on the other hand, are partially absorbed by shifts in the optical depth to reionization $\tau_\mathrm{reio}$ and the primordial spectrum amplitude $A_s$, leaving behind nontrivial modifications to the TT and EE spectra.
Where these modifications are oscillatory, the frequency is set by the BAO scale.
The injection at $1+z = 1100$ shows mild correlation with several cosmological parameters, but the residual also generally retains BAO-frequency oscillations.
Finally, the EDE signals at $1+z > 10^2$ can be partially absorbed by shifts in the scalar spectral index $n_s$, the baryon density $\Omega_b$, and the Hubble parameter $H_0$, as is expected since these models are motivated through their impact on the Hubble tension.

Comparing the various types of signals, we can see key distinguishing features between the primordial and post-inflationary models.
For the post-inflationary features studied here, the dominant effects are an overall change in amplitude and a phase shift, which appear as approximately scale-independent fractional residuals or oscillatory residuals at the BAO frequency, respectively.
Only the oscillations are qualitatively robust to refitting the exotic signals with the $\Lambda$CDM parameters.
On the other hand, while primordial features also give oscillatory residuals, the frequency of the residuals is generally a free parameter.
Hence, if one detects significant residuals in the CMB power spectrum with respect to $\Lambda$CDM, it should be possible to attribute the residual to either primordial or post-inflationary physics.
The only ambiguous case is if the residuals are consistent with oscillations of BAO frequency; however, in this case, the residuals are more likely to have been caused by post-inflationary physics, as one would have to tune the frequency of primordial features to be coincident with the BAOs. 
Furthermore, even when the primordial feature frequency coincides with the BAOs, the transfer functions modulate the amplitude, reducing the degeneracy.
This confirms the intuitive results developed in Sections~\ref{sec:primordial} and ~\ref{sec:late}.

\section{Principal component analysis}
\label{sec:PCA}

The effects of energy injection from different DM models on the CMB power spectrum tend to be highly correlated, and hence are characterized by a small number of parameters---we would like to understand if the same is true of the classes of primordial and EDE-like features studied here. 
PCA offers a technique for identifying the most impactful parameters by finding a basis of principal components (PCs) with orthogonal effects on CMB anisotropies.

Intuitively, PCA systematically identifies which combinations of the basis models produce the most distinguishable imprints on the CMB. 
The first PC captures the direction in model space that leaves the largest, most constrainable signal; subsequent PCs represent orthogonal and progressively less constrained effects. 
If only a few PCs capture most of the ``constraining power'', this tells us that despite having many individual basis models, their effects on the CMB are highly correlated and effectively described by a small number of parameters.
Exotic cosmologies whose effects are mainly captured by these primordial and post-inflationary features can then be expressed in terms of these PCs. 

Beyond identifying these effective parameters, PCA allows us to quantify how degenerate the features are in terms of their effects on the observable. 
By examining correlations between PCs from different model classes, we can assess whether primordial and post-inflationary signals are distinguishable or degenerate.

These methods are described in greater detail in Refs.~\cite{Finkbeiner:2011dx,Slatyer:2016qyl,Xu:2024vdn} and the main points are summarized in Section~\ref{sec:PCA_methods}.
Readers primarily interested in scientific results may skip directly to Section~\ref{sec:PCA_results}.

\subsection{Methods}
\label{sec:PCA_methods}

The basis models should be chosen such that the signals we are interested in can be represented as linear combinations of the basis models.
Our basis will consist of the primordial features described in Section.~\ref{sec:primordial} with vanishing phase, $\phi = 0$, which are characterized by $k_0$ for the sharp features and $\omega/H$ for the resonant features; exotic energy injections with decay/annihilation-like redshift dependence; and transient EDE with varying $z_\mathrm{EDE}$ and $f_\mathrm{EDE}$, but parameters otherwise fixed as described in Section~\ref{sec:late}.
We will discretize the $p(z)$ exotic injection functions as sums of injections that are narrow in time, $p(z) = \sum \alpha_i G_i(z)$, where the $G_i(z)$ basis functions are log-normal functions centered at $z_i$ with a width of $\sigma = \Delta \log z$, where $\Delta \log z$ is the spacing between $z_i$, i.e.
\begin{gather}
    G_i(z) \propto \exp\!\left[ -\frac{\left(\log_{10} (1+z) - \log_{10} (1+z_\text{i})\right)^2}{2\sigma^2} \right].
\end{gather}
The functions are normalized such that $\int d\log z \, G_i (z) = \Delta \log z$.
We take 40 logarithmically spaced redshift bins between $80 \leq z \leq 1300$ for annihilations and 40 logarithmically spaced redshift bins between $10 \leq z \leq 1300$ for decays.

We conduct six analyses, by assuming any primordial features are either of the resonant \textit{or} sharp form and that post-inflationary features resemble either decaying dark matter, annihilating dark matter, or and an EDE-like component.
For each basis model, which we denote with $i \in [1,N]$, and each multipole $\ell$, we calculate the derivative of the CMB power spectrum with respect to the amplitude of the basis model, $\partial (\Delta C^{XX}_\ell)_i / \partial A_i$, where $XX \in \{TT, TE, EE\}$ and $A_i \in \{A_{\rm lin}, A_{\rm log}, \alpha_i, f_\mathrm{EDE}\}$.

From the derivatives, we then calculate the $n_\ell \times N$ transfer matrices (where $n_\ell$ denotes the number of multipoles we include), which map exotic features into perturbations of the CMB anisotropy spectrum. 
\begin{equation}
    (T_e)_{\ell i} = \left( \frac{\partial (\Delta C^{TT}_\ell)_i}{\partial A_i}, \frac{\partial (\Delta C^{EE}_\ell)_i}{\partial A_i}, \frac{\partial (\Delta C^{TE}_\ell)_i}{\partial A_i} \right),
    \label{eqn:Tli}
\end{equation}
In other words, $(T_e)_{\ell i}$ is a three-component vector containing the derivatives of the $TT$, $EE$, and $TE$ spectra at a certain $\ell$ for some model labeled by $i$. 
With the transfer matrix, we construct the Fisher matrix for exotic model parameters, $F_e$, as follows:
\begin{equation}
    (F_e)_{ij} = \sum\limits_{\ell} (T_e)_{\ell i}^T \cdot \Sigma_\mathrm{cov}^{-1} \cdot (T_e)_{\ell j}. 
\end{equation}
$F_e$ is an $N \times N$ matrix and $\Sigma_\mathrm{cov}$ is the covariance matrix (see e.g. Refs.~\cite{Tegmark:1996bz,Jungman:1995bz,Verde:2009tu}):
\footnote{There is a typo in some previous works with this covariance matrix, where the factor of 1/2 in the last entry is missing~\cite{Finkbeiner:2011dx,Slatyer:2016qyl,Xu:2024vdn}.}
\begin{equation}
    \Sigma_\mathrm{cov} = \frac{2}{2\ell + 1}
    \begin{pmatrix}
        (C_\ell^{TT})^2 & (C_\ell^{TE})^2 & (C_\ell^{TT} C_\ell^{TE}) \\
        (C_\ell^{TE})^2 & (C_\ell^{EE})^2 & (C_\ell^{EE} C_\ell^{TE}) \\
        (C_\ell^{TT} C_\ell^{TE}) & (C_\ell^{EE} C_\ell^{TE}) & \frac12 [ (C_\ell^{TE})^2 + C_\ell^{TT} C_\ell^{EE} ] \\
    \end{pmatrix}.
\end{equation}
To include experimental noise one can replace the power spectrum with $C_\ell^{TT,EE} \rightarrow C_\ell^{TT,EE} + N_\ell^{TT,EE}$, where \begin{equation}
    N_\ell = (\omega_p)^{-1} e^{\ell(\ell+1) \theta^2}.
\end{equation} 
Here, $N_\ell$ is the effective noise power spectrum, $\theta$ is the beam width (FWHM = $\theta \sqrt{8 \ln 2}$), and $(\omega_p)^{-1} = (\Delta T \times \mathrm{FWHM}) ^2$ is the raw beam sensitivity. 
To include the effect of partial sky coverage, we divide $\Sigma_\mathrm{cov}$ by $f_\mathrm{sky} = 0.65$. 
In this study, we consider \textit{Planck}~\cite{Planck:2018vyg} and an experiment that is cosmic variance limited (CVL) in temperature and polarization up to $\ell = 2500$. 
Table \ref{tab:noise} gives the beam width and sensitivity parameters, which are the same as were used in Refs.~\cite{Xu:2024vdn,Finkbeiner:2011dx}.

\begin{table}
    \centering
    \begin{tabular}{c c c c}
    \hline
     Experiment & FWHM [arcmin] & $10^6 \Delta T/T$ (I) & $10^6 \Delta T/T$ (Q, U) \\ 
     \hline
     Planck (100 GHz) & 10 & 2.5 & 4.0 \\  
     Planck (143 GHz) & 7.1 & 2.2 & 4.2 \\
     Planck (217 GHz) & 5.0 & 4.8 & 9.8 \\
     \hline
    \end{tabular}
    \caption{Parameters characterizing the noise power spectrum for \textit{Planck}.}
    \label{tab:noise}
\end{table}

We marginalize over the cosmological parameters, which are the physical baryon density $\omega_b = \Omega_b h^2$, the physical CDM density $\omega_c = \Omega_c h^2$, the
primordial scalar spectral index $n_s$, the normalization $A_s$, the redshift of reionization $z_\mathrm{re}$, and the Hubble parameter $H_0$. 
Similar to what was done for exotic features, we calculate the effect of varying the cosmological parameters on $\Delta C_\ell$, and construct their transfer matrix, $(T_c)_{\ell i}$, and Fisher matrix, $F_c$.
The full Fisher matrix is then given by \begin{equation}
    F_0 = \begin{pmatrix}
        F_e & F_v \\
        F_v^T & F_c
    \end{pmatrix},
\end{equation}
where $F_v$ contains cross terms, 
\begin{equation}
    (F_v)_{ij} = \sum\limits_{\ell} (T_e)_{\ell i}^T \cdot \Sigma_\mathrm{cov}^{-1} \cdot (T_c)_{\ell j}. 
\end{equation} 
The marginalized Fisher matrix can be calculated with block-matrix inversion and is given by $F = F_e - F_vF_c^{-1}F_v^T$. 

The basis of PCs is then derived by diagonalizing the marginalized Fisher matrix
\begin{equation}
    F = W^T \Lambda W, \,\, \Lambda = \mathrm{diag} (\lambda_1, \lambda_2, \dots, \lambda_N) .
\end{equation} 
To be explicit, we will use $ijk$ indices to indicate when we are using the model basis and $abc$ indices to indicate when we are using the PC basis.
Then, the $a$-th PC is then given by the $a$-th row of $W$, which is the eigenvector $\vec e_a$ with eigenvalue $\lambda_a$. 
By convention, the eigenvectors are normalized such that they are orthonormal and ordered by decreasing eigenvalue, so $\lambda_1$ is the largest eigenvalue. 

A key assumption of the PCA is that the change in the observable, $(\Delta C_\ell)_i$, is linear with respect to the amplitude of the basis models; by calculating the derivatives using a polynomial fit, we can also test the degree of nonlinearity.
We have verified that the nonlinearity in the variation of the $C_\ell$'s with respect to the cosmological parameters is less than $2\%$, sub-percent level for the inflationary features, $\lesssim 10\%$ for post-recombination energy injection, and sub-percent level for the EDE-like features.
In addition, since we are primarily interested in studying the distinguishability of different classes of models through the shape of signal, we need to ensure that the amplitude of signals from one class of models is not being normalized very differently from another.
In other words, if $\alpha_i$ is normalized very differently (e.g. has different units) compared to $A_{\rm lin}$ or $A_{\rm log}$, then the $\partial (\Delta C^{XX}_\ell)_i / \partial A_i$ derivatives will also have very different normalization between classes of models, and the models with the larger derivatives will dominate the Fisher matrix and therefore the first several PCs.
To mitigate this issue, we multiply the exotic energy injection derivatives by arbitrary constants to ensure that the contributions from primordial features and post-inflationary features have a similar amplitude.
We find that multiplying the energy injection derivatives by a factor of $2 \times \langle |T_\mathrm{ann/dec}| \rangle / \langle |T_\mathrm{sharp}| \rangle$, where $T_\mathrm{ann/dec}$ is the transfer matrix for annihilation or decay basis models and $T_\mathrm{sharp}$ the transfer matrix for sharp features, results in a well-behaved PCA.

To estimate the ability of an experiment to constrain an exotic model, we decompose the model in terms of the basis signals as $M = \sum_{i=1}^N \alpha_i M_i$, where $M_i$ denotes the basis models and $\vec \alpha = \{\alpha_i\}$ is a set of coefficients.  
Then the 2$\sigma$ sensitivity estimate on the amplitude of the model, $A_M$, corresponds to 
\begin{equation}
    A_M < \frac{2}{\sqrt{\sum_{a=1}^{N_\mathrm{max}} (\vec \alpha \cdot \vec e_a)^2 \lambda_a}}.
\end{equation}
For the basis model corresponding to the index $i$ with coefficients $\alpha_j = \delta_{ij}$, this corresponds to a sensitivity estimate of
\begin{equation}
    A_i < \frac{2}{\sqrt{\sum_{a=1}^{N_\mathrm{max}} (\vec e_a)_i^2 \lambda_a}} .
    \label{eqn:PCA_limit}
\end{equation}
However, most of the principal components are very subdominant compared to the first, and the limit can be well approximated with a handful of PCs; this was the case for exotic energy injection, where the constraint on annihilating and decaying dark matter was dominated by the first PC~\cite{Finkbeiner:2011dx,Slatyer:2016qyl}.
Since one can think of the sum of the eigenvalues of the PCs in some sense as the total constraining power of the CMB anisotropies on exotic models (or more precisely, the total signal variance), we will only keep principal components that contribute at least 1\% of the total variance---for each analysis, this corresponds to keeping 11 to 14 PCs.

In Appendix~\ref{app:PCA_injection}, we derive sensitivity estimates for primordial features and exotic energy injection in Planck using only the most significant PCs.
This also serves as validation of our PCA, since we find the exotic energy injection results (in particular, the values of $z_\mathrm{inj}$ to which the CMB power spectrum is most sensitive) to be broadly consistent with the results of Ref.~\cite{Finkbeiner:2011dx}.

\subsection{Results}
\label{sec:PCA_results}

\begin{figure*}
    \centering
    \includegraphics[width=\textwidth]{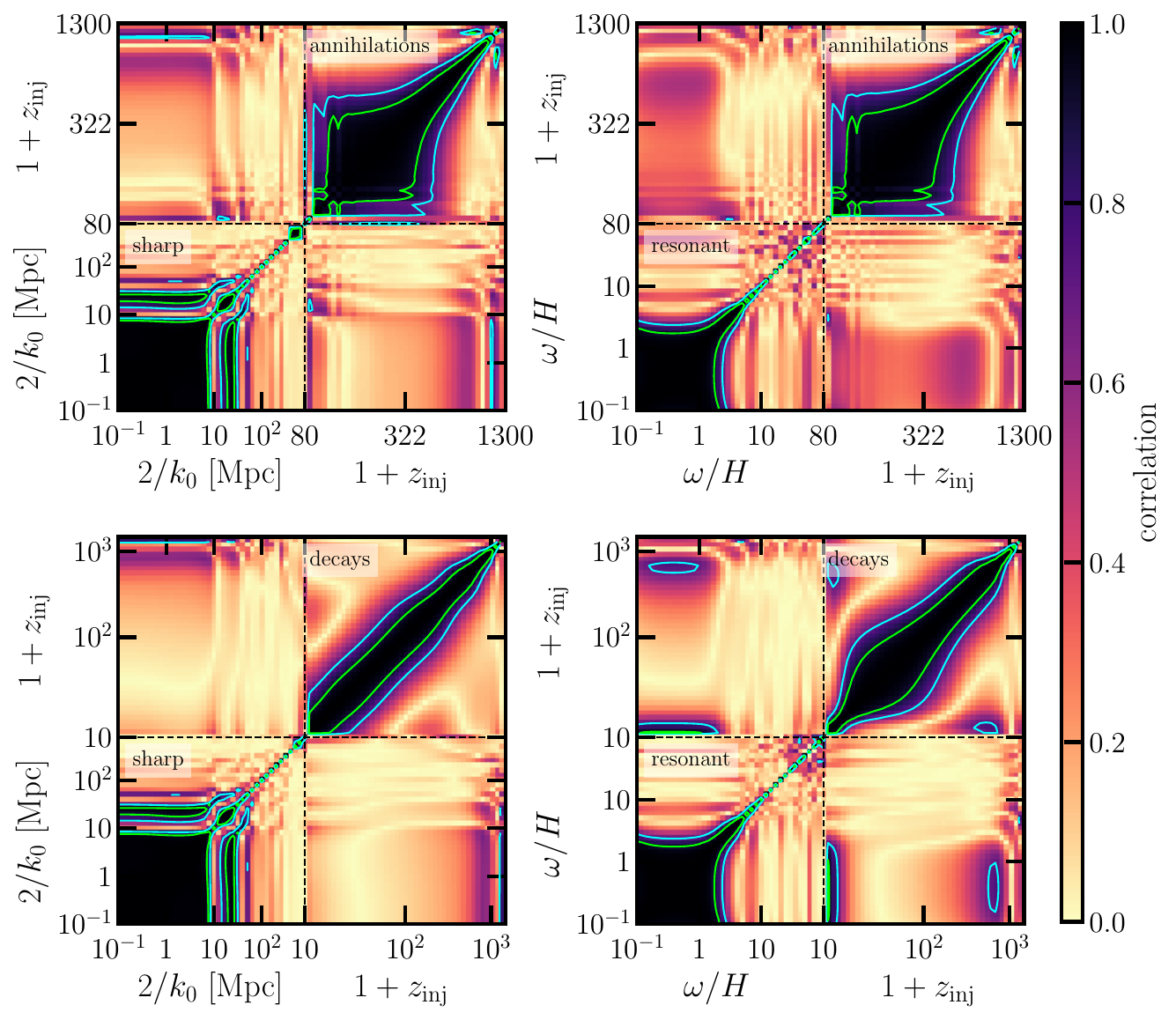}
    \caption{
    Correlations between primordial features and exotic energy injection, when truncating the PCs to those contributing $>1\%$ to the constraining power and after normalizing the vectors to unity.
    }
    \label{fig:corr-normalized}
\end{figure*}
\begin{figure*}
    \centering
    \includegraphics[width=\textwidth]{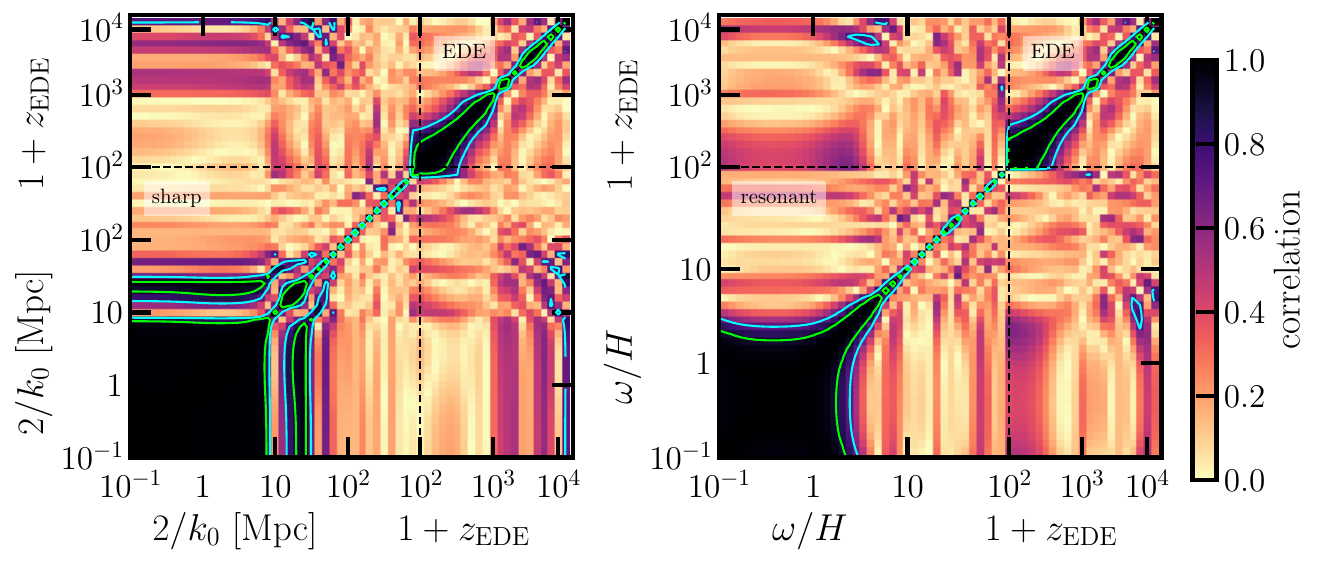}
    \caption{
    Correlations between primordial features and EDE, when truncating the PCs to those contributing $>1\%$ to the constraining power and after normalizing the vectors to unity.
    }
    \label{fig:corr-normalized-EDE}
\end{figure*}
To determine how distinguishable the different classes of models are, we can look at the correlation between the PC coefficients for the basis models.
Generally, one can quantify the degeneracy between basis models $i$ and $j$ through the normalized Fisher matrix,
\begin{equation}
    C_{ij} = \frac{F_{ij}}{\sqrt{F_{ii}\,F_{jj}}},
\end{equation}
where $F$ is the joint Fisher matrix of all basis models, marginalized over the $\Lambda$CDM parameters. 
Values of $|C_{ij}|$ close to unity indicate that the two models produce nearly indistinguishable changes in the CMB spectra. 
When the analysis is restricted to the first $N$ principal components, we replace $F$ by its truncated reconstruction,
\begin{equation}
    F^{(N)}_{ij} = \sum_{a=1}^{N} \lambda_a\, e_{a,i}\, e_{a,j},
\end{equation}
and define the correlation $C^{(N)}_{ij}$ analogously. 
As before, we truncate the PCs to include only those with significant eigenvalues, e.g. contributing $>1\%$ to the total variance.
The correlation between the primordial and exotic injection basis models is shown in Fig.~\ref{fig:corr-normalized}, while the correlation between primordial and EDE-like basis models is shown in Fig.~\ref{fig:corr-normalized-EDE}.
Correlation at the level of 0.7 is shown in bright blue contours and 0.9 in bright green.

Inflationary features with very slow oscillation frequencies all have fairly degenerate effects on the CMB, as can be seen from the dark blob in the lower left of each panel in Figs~\ref{fig:corr-normalized} and \ref{fig:corr-normalized-EDE}.
Intuitively, this is sensible, since as one moves to lower frequencies, the CMB power spectrum residuals contain few wavelengths across the observable range of $\ell$, hence the frequency of the residuals becomes more difficult to measure accurately.
By contrast, at larger frequencies, the effect of the exotic models on the CMB is quite orthogonal from one basis model to the next, as is evident from the lack of cross-correlation between adjacent models.
One can improve the distinguishability of the low-frequency basis models by including more PCs.
Hence, due to the fact that the residual signal in the CMB power spectrum from primordial features can have any frequency, they cannot be captured with a simple set of templates.

Compare this to post-inflationary features.
We find that exotic energy injections at similar redshifts are highly degenerate, as can be seen from the thick band of dark colors around the diagonal in the upper right quadrant of each panel of Fig.~\ref{fig:corr-normalized}.
As was discussed in previous sections, this is due to the fact that exotic energy injection can only cause signals that appear as an overall damping of the CMB power spectrum or a phase shift; this is also consistent with previous findings that exotic energy injection can be described by a very small number of PCs~\cite{Finkbeiner:2011dx,Slatyer:2016qyl}.

Between primordial and exotic injection features, the only degeneracies appear to be mild cross-correlations with the resonant features.
The only correlations greater than 0.9 appear between resonant features and decay-like injections at $1+z_{\rm inj} = 10$, and these mild degeneracies occur where the resonant features have quite long wavelengths.
However, these cross-correlations also decrease if once chooses to include more PCs in the analysis.

With regards to the EDE-like features in Fig.~\ref{fig:corr-normalized-EDE}, we find that fluids transitioning at different values of $z_\mathrm{EDE}$ leave distinct imprints on the CMB and the resulting signals are not very degenerate except for $z_\mathrm{EDE} \lesssim 10^3$.
At this lower redshift range, the changes to the power spectrum are not only similar to each other in shape, but also smaller in amplitude at fixed $f_\mathrm{EDE}$ compared to the cases where $10^3 < z_\mathrm{EDE} < 10^4$---hence, when truncating the PC below some eigenvalue threshold, we begin to lose information for the EDE basis models at low redshifts first. 
When analyzing the primordial and EDE-like features together, we find that there are no models that show cross-correlations greater than 0.9, indicating that the two classes of features are not very degenerate with each other.

\section{Full likelihood analysis with Planck data}
\label{sec:MCMC}

Our PCA analysis in Section~\ref{sec:PCA} strongly suggests that the characteristic oscillatory residuals in the CMB anisotropies produced by primordial features are not degenerate with the corresponding residuals induced by exotic energy injections or motivated EDE models in the post-inflationary era. 
Indeed, the latter sets of models either suppress the CMB power or shift the acoustic peaks, and cannot reproduce the peculiar running of primordial feature oscillations.

In this Section, we want to confirm the results of the PCA analysis by means of a full likelihood analysis performed on real data. While informative and useful for building intuition, the PCA analysis of the previous Section relies on some simplifying assumptions, such as simplified noise modeling, the linearization of the model response, and the Gaussian approximation to the likelihood around the fiducial cosmology. On the other hand, these assumptions may not be satisfied in realistic scenarios, especially for models of primordial features, which are well known to produce multimodal structures in the posterior distributions of their parameters, mainly because different oscillation frequencies can generate multiple isolated maxima in the likelihood.

In passing, our analysis also delivers updated constraints using the most recent Planck data release on exotic energy injections.
In fact, to the best of our knowledge, besides the PCA sensitivity estimate of Ref.~\cite{Finkbeiner:2011dx}, the simplified log-normal injections considered in this paper have never been constrained in the literature, and our results therefore provide the {\em first} constraints on these models. 
We stress, however, that we do not claim that this shape of the energy injection is realistic, and we only consider it for illustrative purposes, as well as for the first joint constraints on primordial features and exotic energy injections of the type considered here.

\subsection{Datasets and settings of the analysis} 

As in Section~\ref{sec:PCA}, the observable considered in the analysis is the angular power spectrum of CMB anisotropies. We employ TT, EE, and TE measurements from the Planck satellite. While other ground-based experiments, such as ACT~\cite{ACT:2025fju} and SPT~\cite{SPT-3G:2025bzu}, have recently delivered high-quality measurements at smaller scales than Planck, we focus on the latter for our analysis, as these data cover enough multipoles to probe the point anticipated by our PCA analysis. Our goal here is not to derive the strongest possible constraints on either class of models, but rather to test whether their characteristic signatures become degenerate in a full likelihood analysis.

We use low-$\ell$ data from the Commander (temperature) and SRoll2 (E-mode polarization) large-scale anisotropy likelihoods~\cite{Delouis:2019SRoll2,Pagano_2020}. We combine these data with high-$\ell$ data. For the latter, we consider the most recent CamSpec likelihood~\cite{Efstathiou:2019mdh,Rosenberg}, based on the Planck PR4 NPIPE maps~\cite{Planck:2020olo}. These maps have lower noise and improved control of several systematic effects compared to previous Planck releases, making them particularly well suited for the present analysis.

In order to explore the degeneracies between model parameters, we perform the following set of analyses:
\begin{itemize}
    \item {\bf $\Lambda$CDM.} This is our baseline with which we compare the results of the extended models. We vary the cosmological parameters $\{A_s,\,n_s,\,\omega_c,\,\omega_b,\,H_0,\,z_{\rm reio}\}$. In particular, we use $z_{\rm reio}$ instead of $\tau_{\rm reio}$ in our analysis, since using the former as an input for CLASS is more stable when including exotic energy injection. 
    All the standard cosmological parameters are well constrained, and we adopt the same broad priors also in the analyses of the extended models.
    
    \item {\bf Sharp feature.} We vary all the feature parameters using the following priors: $A_{\rm lin}\in[0,1]$, $\log_{10}\omega_{\rm lin}\in[0,\,2.1]$, and $\phi_{\rm lin}\in[0,\,2\pi]$, where $\omega_{\rm lin}\equiv 2/k_0$. We note that a logarithmic prior on $\omega_{\rm lin}$ is this equivalent to a logarithmic prior on $k_0$.
    
    \item {\bf Resonant feature.} We vary all the feature parameters using the following priors: $A_{\rm log}\in[0,1]$, $\log_{10}\omega_{\rm log}\in[0,\,2]$, and $\phi_{\rm log}\in[0,\,2\pi]$, where we identified $\omega_{\rm log}\equiv\omega/H$.

    \item {\bf Exotic Energy Injection.} We are interested in narrow energy injections, such as those analyzed in the previous Section. We therefore fix $\sigma_{\rm inj}=0.02$ in our analysis. 
    We then adopt a logarithmic prior on the effective lifetime, $\log_{10} (\Gamma / 1 \,\mathrm{s}^{-1})\in[-26.5,\,-21.5]$, which is sufficiently broad to encompass the region allowed by current $y$-distortion limits, and a logarithmic prior $\log_{10} z_{\rm inj}\in[0.6,\,3.15]$ to allow the energy injection to occur all the way from reionization to redshifts larger than recombination. 

    \item  {\bf Early Dark Energy.} Also in this case, we are interested in narrow modifications to the expansion of the Universe. We therefore fix $\sigma_{\rm EDE}=0.1$ and $w_f=7$ in our analysis. We assume a constant speed of sound $c_s^2=1$. As most of the observable effects in EDE occur for $z_{\rm EDE}\sim z_{\rm rec}$, we follow earlier works~\cite{Lin:2019qug,Smith:2022hwi} and adopt the prior $\log_{10}z_{\rm EDE}\in[2.9,\,4]$. We set a very broad prior on the amplitude of the injection $f_{\rm EDE}\in[0,\,0.95]$.
    
    \item {\bf Exotic and/or Early Dark Energy Injection + primordial feature.} Finally, we perform two joint analyses per post-inflationary model, combining them with either the sharp or the resonant primordial-feature template. We use the priors on the extra parameters reported above.
\end{itemize}

\begin{table*}[t]
\centering
\caption{Best-fit parameter values. The $\Delta\chi^2_{\rm CMB}$ is defined as $\chi^2_{{\rm CMB},{\rm model}}-\chi^2_{{\rm CMB},\Lambda{\rm CDM}}$, so negative values correspond to an improved fit.}
\label{tab:bestfit_parameters}
\small

    \scriptsize
    \setlength{\tabcolsep}{3.5pt}
    \renewcommand{\arraystretch}{0.92}
    
\begin{tabular}{@{}lcccccccc@{}}
\hline\hline
Parameter & Inj & EDE & Res & Res-inj & Res-EDE & Sharp & Sharp-inj & Sharp-EDE \\
\hline
     &&  && & &&& \\
    {Varied} &&&&&  && &  \\
     &&  && &&&&  \\
    $A_s\times10^{9}$ & $2.0686$ & $2.0613$ & $2.0343$ & $2.0606$ & $2.0735$ & $2.0579$ & $2.0668$ & $2.0573$ \\
$n_s$ & 0.96445 & 0.96355 & 0.96253 & 0.96380 & 0.96306 & 0.96076 & 0.96118 & 0.96077 \\
$H_0$ & 67.42 & 67.48 & 67.55 & 67.26 & 67.39 & 67.08 & 66.94 & 67.28 \\
$\omega_b$ & 0.02219 & 0.02221 & 0.02223 & 0.02222 & 0.02227 & 0.02209 & 0.02219 & 0.02216 \\
$\omega_c$ & 0.11836 & 0.11922 & 0.11906 & 0.11977 & 0.12065 & 0.11994 & 0.12054 & 0.11995 \\
$z_{\rm reio}$ & 7.50 & 7.09 & 6.53 & 6.77 & 6.89 & 6.97 & 7.07 & 7.09 \\
$\Gamma\times10^{22}$ & 0.50 & -- & -- & 1.63 & -- & -- & 0.06 & -- \\
$\log_{10} z_{\rm inj}$ & 3.06 & -- & -- & 3.26 & -- & -- & 3.26 & -- \\
$f_{\rm EDE}$ & -- & 0.0006 & -- & -- & 0.0111 & -- & -- & 0.0102 \\
$\log_{10} z_c$ & -- & 3.42 & -- & -- & 3.22 & -- & -- & 3.53 \\
$A_{\rm log}$ & -- & -- & 0.0270 & 0.0232 & 0.0221 & -- & -- & -- \\
$\phi_{\rm log}$ & -- & -- & 5.28 & 4.77 & 4.43 & -- & -- & -- \\
$\log_{10}\omega_{\rm log}$ & -- & -- & 1.70 & 1.71 & 1.71 & -- & -- & -- \\
$A_{\rm lin}$ & -- & -- & -- & -- & -- & 0.0177 & 0.0189 & 0.0249 \\
$\phi_{\rm lin}$ & -- & -- & -- & -- & -- & 5.97 & 4.85 & 5.23 \\
$k_0\times 10^3 [{\rm Mpc}^{-1}]$ & -- & -- & -- & -- & -- & 2.62 & 2.57& 2.60 \\
\hline

     &&  && & &&& \\
    {Derived} &&&&&  && &  \\
     &&  && &&&&  \\$z_{\rm inj}$ & 1154.32 & -- & -- & 1834.82 & -- & -- & 1825.94 & -- \\
$\omega_{\rm lin}$ & -- & -- & -- & -- & -- & 760.52 & 776.83 & 770.28 \\
$\omega_{\rm log}$ & -- & -- & 50.08 & 50.99 & 51.04 & -- & -- & -- \\
$\tau_{\rm reio}$ & 0.05247 & 0.04838 & 0.04329 & 0.04535 & 0.04644 & 0.04692 & 0.04799 & 0.04826 \\
\hline
$\Delta\chi^2_{\rm CMB}$ & -0.8 & -1.1 & -7.9 & -10.5 & -11.3 & -6.9 & -9.8 & -8.0
   \\ 
    $\Delta\ln\mathcal{B}$
    & $0.35\pm0.34$
    & $3.16\pm0.39$
    & $1.74\pm0.39$
    & $2.34\pm0.35$
    & $4.80\pm0.40$
    &$1.37\pm0.39$ & $1.77\pm 0.35$
    & $4.99\pm0.42$
     \\
\hline\hline
\end{tabular}
\end{table*}
\begin{figure*}
    \centering
    \includegraphics[width=\textwidth]{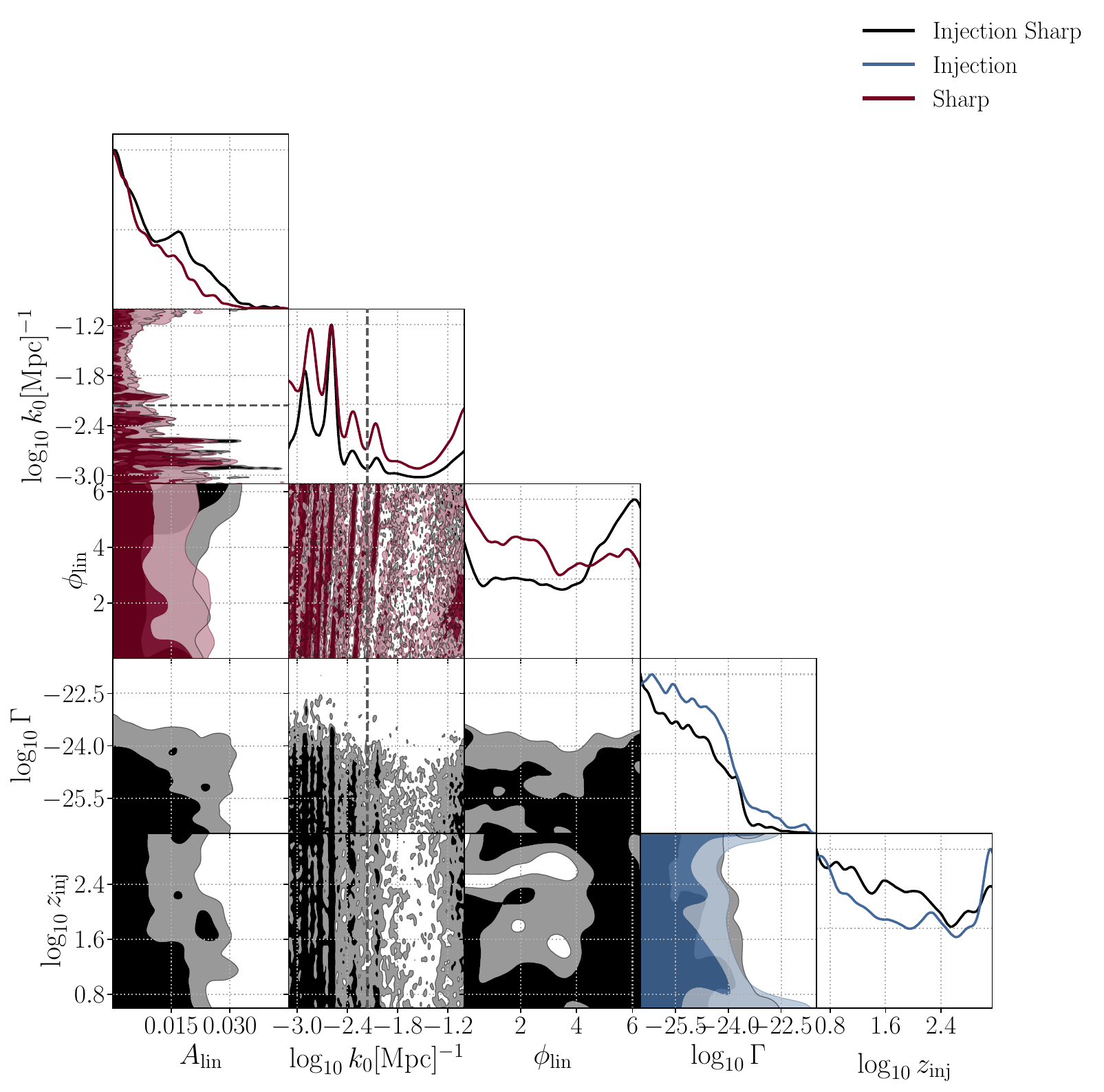}
    \caption{
    1D and 2D posterior distributions of the cosmological parameters in our models. Here we consider $\Lambda$CDM, post-inflationary narrow injection, sharp features and joint  post-inflationary narrow injection plus sharp features. The dataset used in our analysis is CamSpec NPIPE (PR4).
    The thick dashed line represents the BAO frequency $k_0=k_{\rm BAO}$. 
    }
    \label{fig:constraints-sharp}
\end{figure*}

\begin{figure*}
    \centering
    \includegraphics[width=\textwidth]{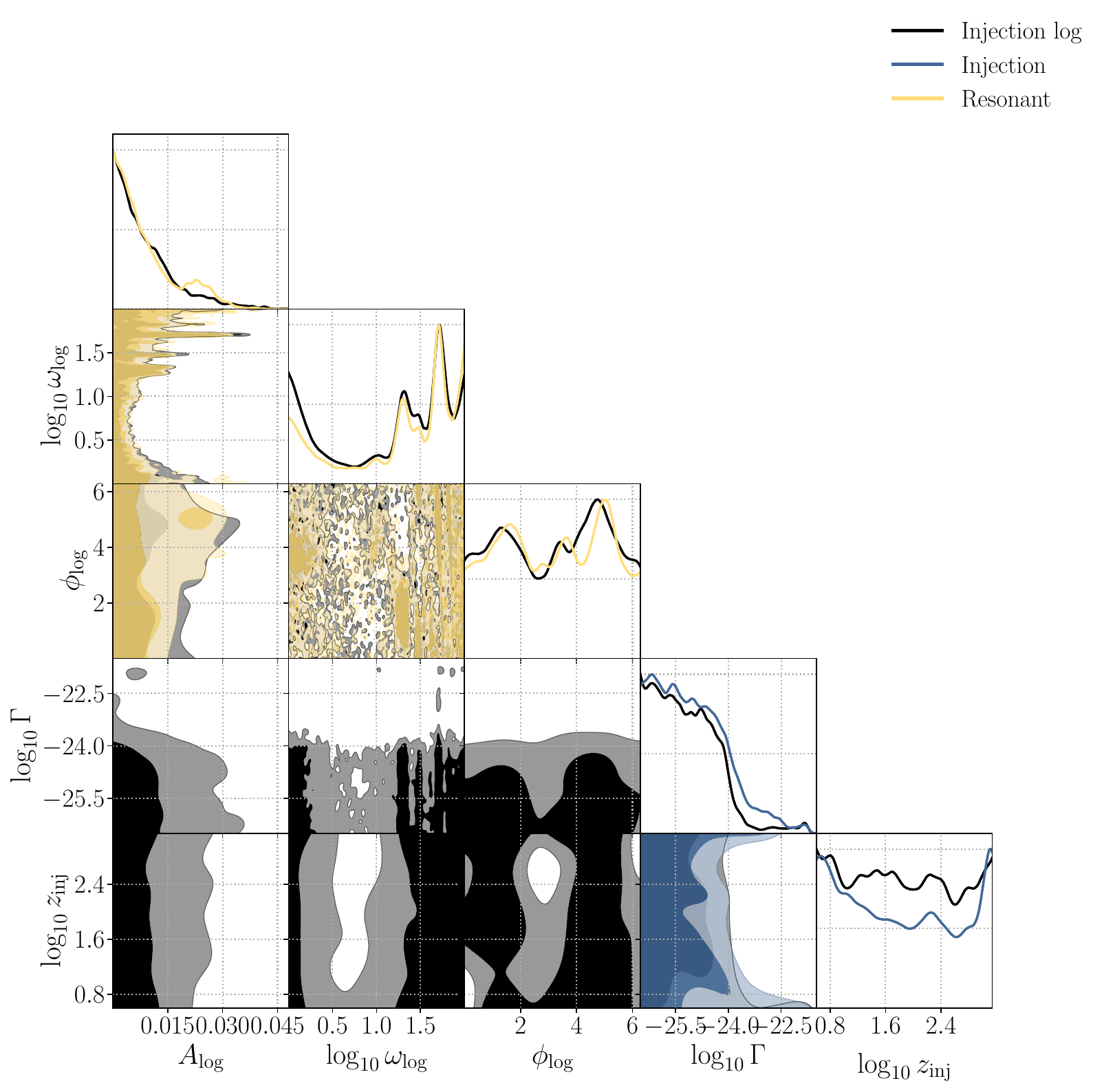}
    \caption{
    1D and 2D posterior distributions of the cosmological parameters in our models. Here we consider $\Lambda$CDM, post-inflationary narrow injection, resonant features and joint  post-inflationary narrow injection plus resonant features. The dataset used in our analysis is CamSpec NPIPE (PR4).
    }
    \label{fig:constraints-log}
\end{figure*}

\begin{figure*}
    \centering
    \includegraphics[width=\textwidth]{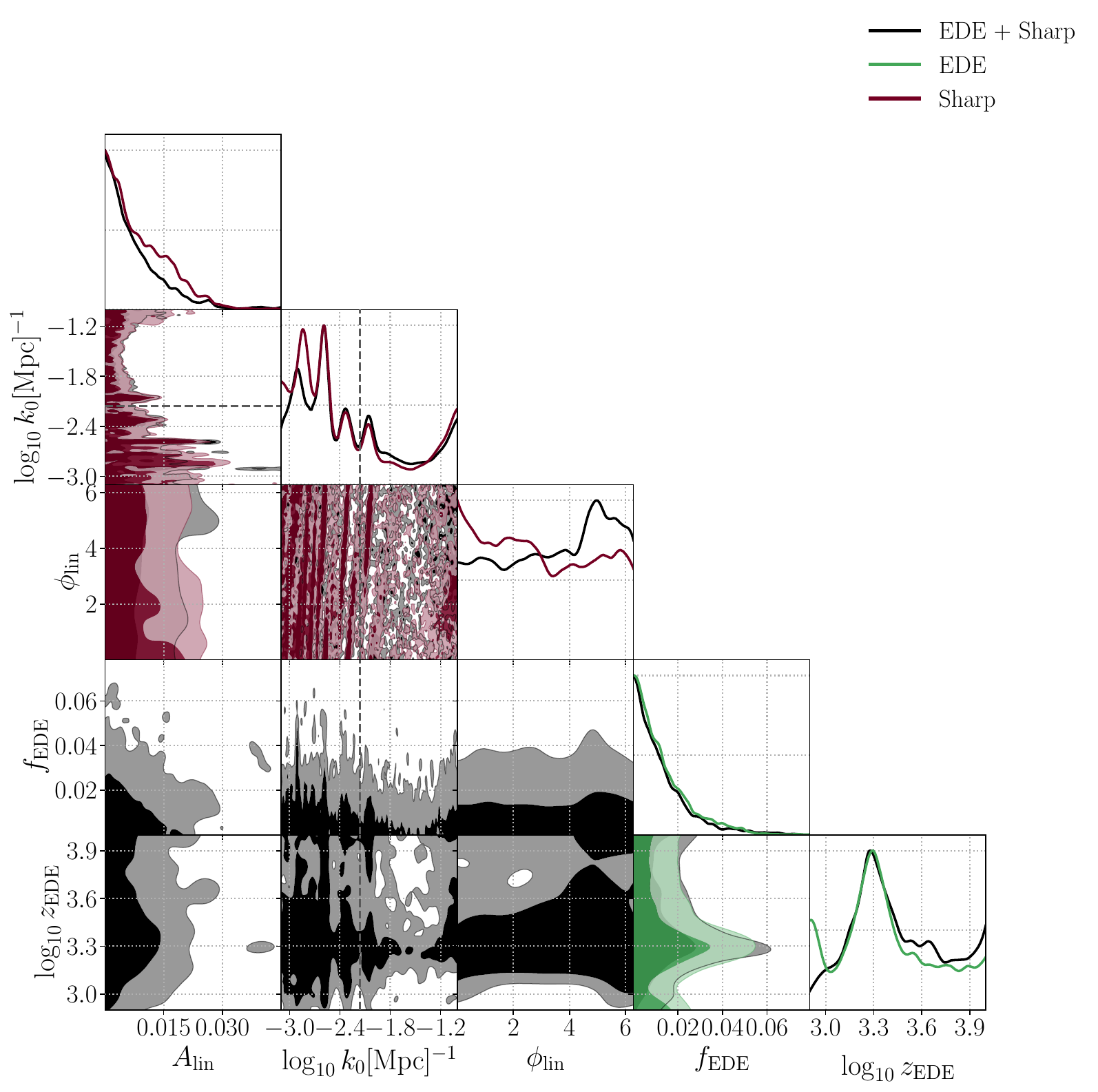}
    \caption{
    1D and 2D posterior distributions of the cosmological parameters in our models. Here we consider $\Lambda$CDM, post-inflationary EDE narrow injection, sharp features and joint  post-inflationary narrow injection plus sharp features. The dataset used in our analysis is CamSpec NPIPE (PR4).  }
    \label{fig:constraints-ede-sharp}
\end{figure*}

\begin{figure*}
    \centering
    \includegraphics[width=\textwidth]{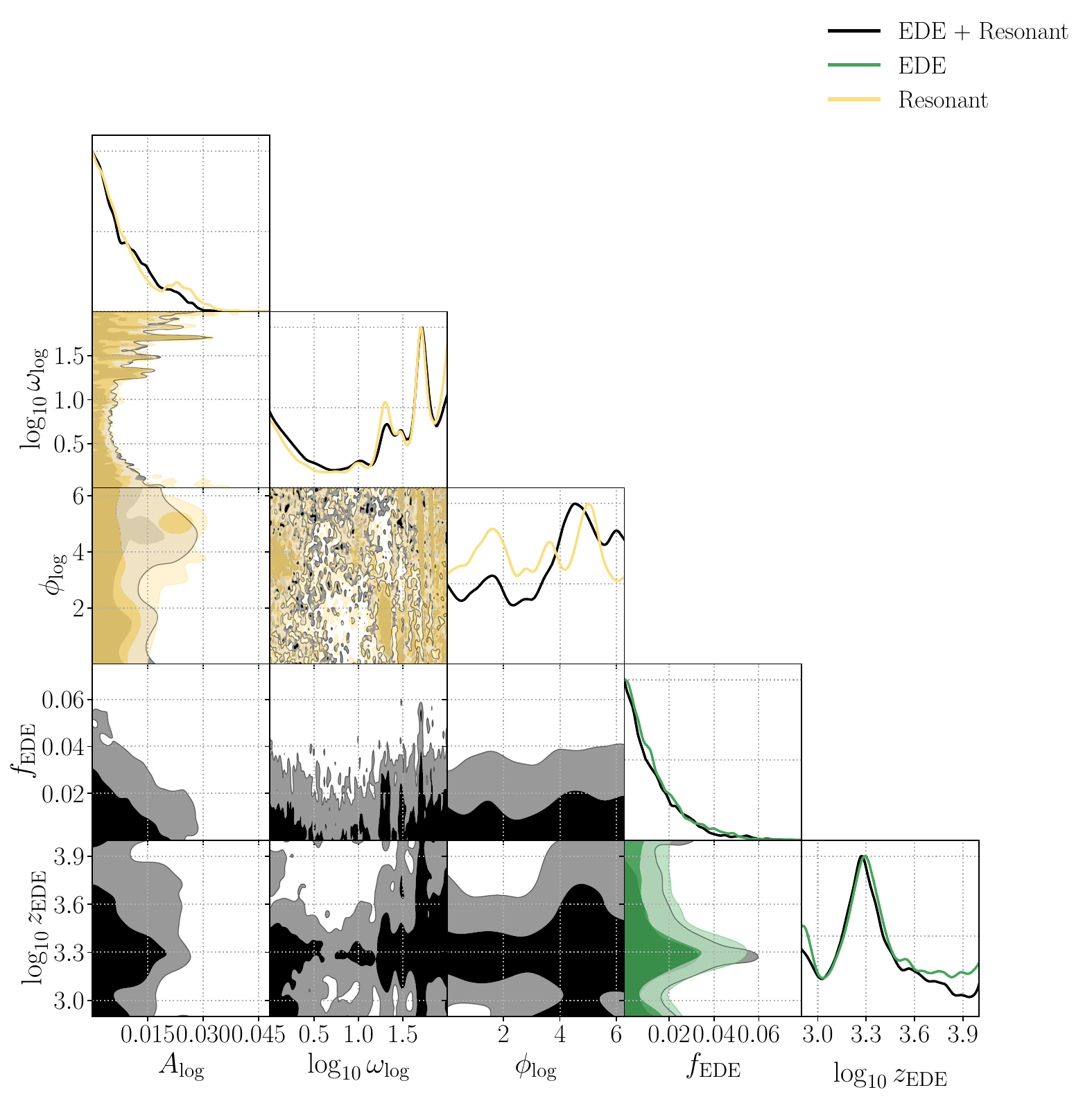}
    \caption{
    1D and 2D posterior distributions of the cosmological parameters in our models. Here we consider $\Lambda$CDM, post-inflationary EDE narrow injection, resonant features and joint  post-inflationary narrow injection plus resonant features. The dataset used in our analysis is CamSpec NPIPE (PR4). 
    }
    \label{fig:constraints-ede-log}
\end{figure*}

\begin{figure*}
    \centering\includegraphics[width=0.9\textwidth]{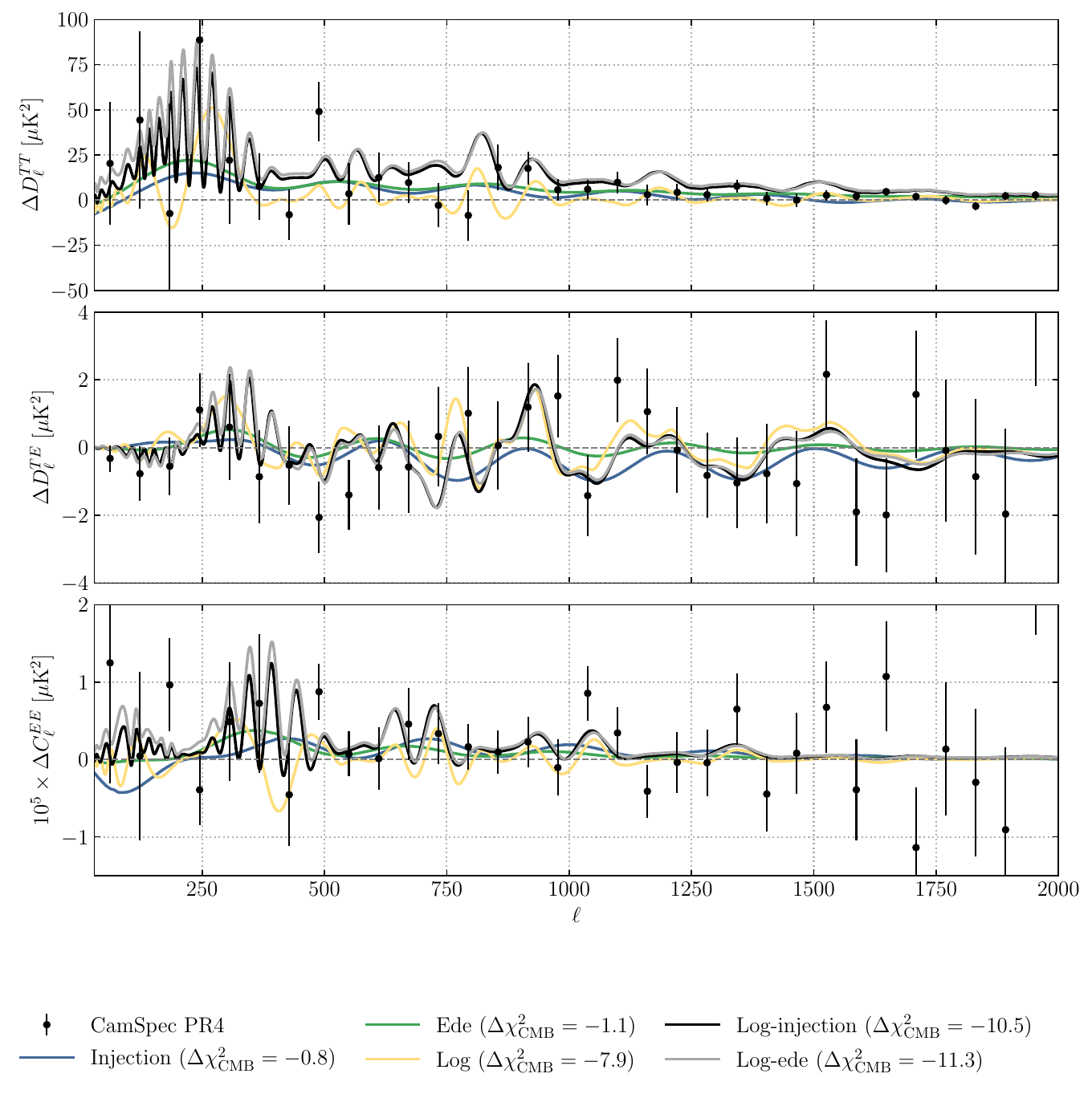}
    \caption{
    CMB residuals for the best fit candidates for $\Lambda$CDM, post-inflationary injections, sharp features and joint  post-inflationary injections plus resonant features. The dataset used in our analysis is CamSpec NPIPE (PR4).
    }
    \label{fig:bestfit-log}
\end{figure*}

\begin{figure*}
    \centering\includegraphics[width=0.9\textwidth]{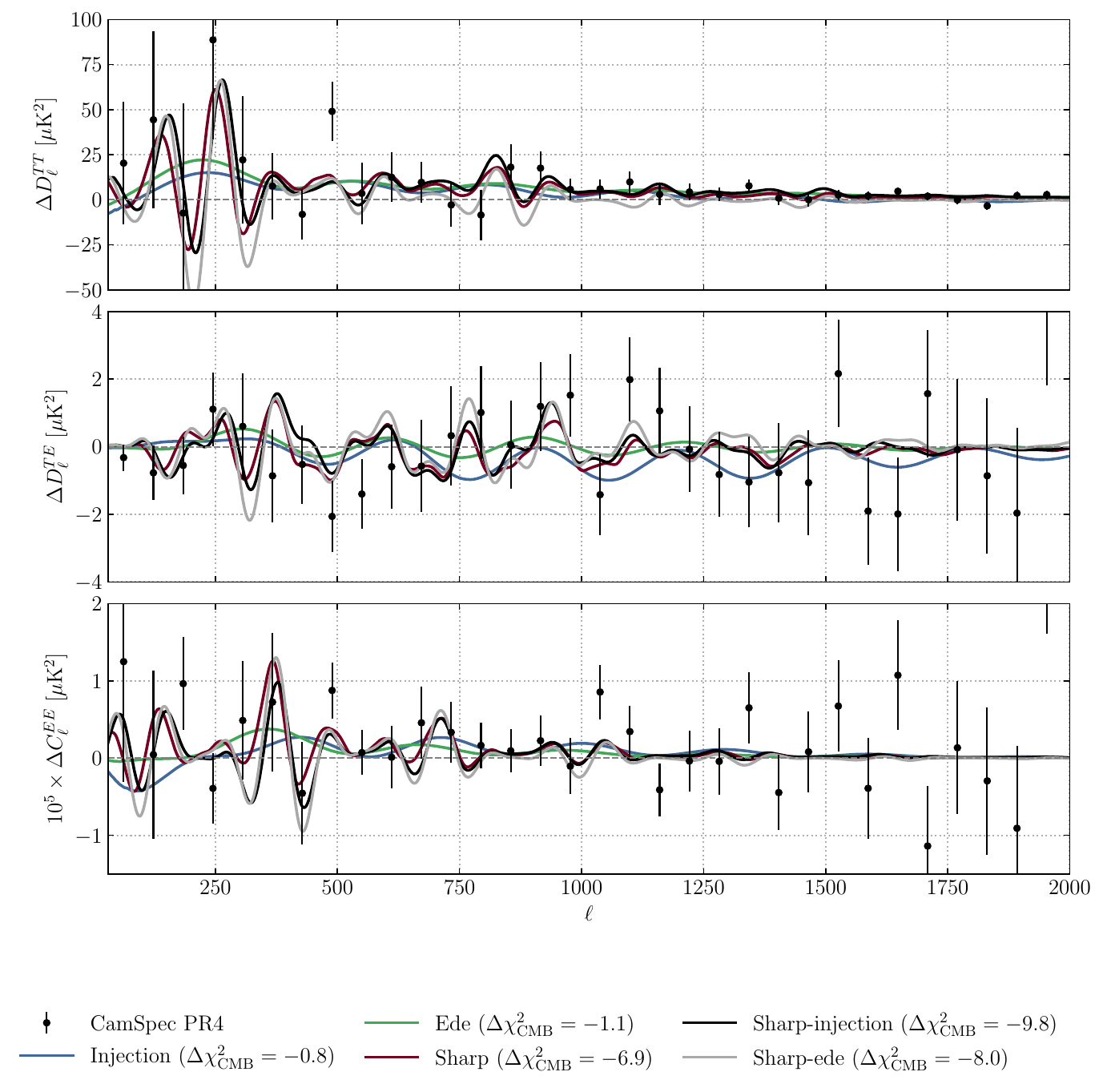}
    \caption{
    CMB residuals for the best fit candidates for $\Lambda$CDM, post-inflationary injections, sharp features and joint  post-inflationary injections plus sharp features. The dataset used in our analysis is CamSpec NPIPE (PR4).
    }
    \label{fig:bestfit-sharp}
\end{figure*}

We sample the Planck likelihood using the nested-sampling algorithm implemented in PolyChord~\cite{Handley_2015,Handley:2015fda} as an add-on to Cobaya~\cite{Torrado:2020dgo}. This also allows us to compute the evidence of the models, which we use to report the natural logarithm of the Bayes factor,
$\ln\mathcal{B}\equiv\ln Z_{\Lambda{\rm CDM}}-\ln Z$, quantifying model selection, so that positive values favor $\Lambda$CDM over the extended model. In all the runs, we employ 1500 live points. As a convergence criterion, we require the estimated remaining contribution to the evidence to satisfy $\Delta\ln Z=0.01$. Nested sampling is particularly useful here, as it captures the multimodal structure of the posterior distribution, which is not easy to explore with techniques such as Markov Chain Monte Carlo.

In addition to this Bayesian estimator, we also compute the best-fit points from the chains and quote the corresponding
$\Delta\chi^2\equiv\chi_{\rm model}^2-\chi^2_{\Lambda{\rm CDM}}$, so that negative values correspond to an improved fit with respect to $\Lambda$CDM. We obtain the best-fit points using the BOBYQA algorithm~\cite{Cartis:2018PyBOBYQA} implemented in Cobaya. The best-fit parameters, the associated $\Delta\chi^2$, and the Bayes factors of the models are quoted in Tab.~\ref{tab:bestfit_parameters}.

We follow the Planck's conventions~\cite{Planck:2018vyg}, and model the Standard Model neutrinos as a combination of one massive species with a mass of 0.06 eV and two massless species, enforcing $N_{\rm eff}=3.046$.
We use {\tt HyRec}~\cite{Ali-Haimoud:2010hou} to compute recombination physics and {\tt HMcode}~\cite{Mead:2020vgs} to calculate the non-linear corrections to the matter power spectrum.

\subsection{Results of the data analysis} 
\label{sec:triangles}

The results of our analyses are reported in Figs.~\ref{fig:constraints-sharp} and~\ref{fig:constraints-log}, where we plot the constraints on the injection parameters together with those of the sharp and resonant feature models, respectively. Before discussing the joint constraints, let us start by commenting on the single-model constraints.

{\bf Single models.} For the primordial features, both the sharp and resonant feature models show no evidence for a detection. This can be seen from the 1D posterior distributions of the amplitudes, $A_{\rm lin}$ and $A_{\rm log}$, which are consistent with upper bounds. While these bounds can be marginalized over the other parameters, it is instructive to inspect how the locally allowed amplitude of the feature depends on the linear or logarithmic frequency, and how several local high-likelihood peaks emerge in the corresponding 2D posterior distributions.
We stress that our data analyses have converged, despite the posterior distributions of the feature parameters appearing highly structured. This is a genuine property of the posterior distributions, which are strongly multimodal and exhibit several local structures.
As can be seen from Tab.~\ref{tab:bestfit_parameters}, the best-fit points correspond to the most prominent local high-likelihood peaks, located at $k_0=2.6\times10^{-3}\,{\rm Mpc}^{-1}$ and $\log_{10}\omega_{\rm log}=1.70$.
We also remark that the previously identified best-fit candidate for the constant-amplitude resonance model at $\omega_{\rm log}\simeq18$, which was robust between PR2 and PR3 data releases (see e.g.~\cite{Akrami:2018odb,Braglia:2021rej,Braglia:2022ftm}), has now significantly less statistical significance with the new PR4 data~\cite{Raffaelli:2026bhk}.

These signals help improve the fit to the CMB residuals, as shown in Figs.~\ref{fig:bestfit-log} and \ref{fig:bestfit-sharp}, with the improvement mostly driven by TT data. We note that our results are in good agreement with the findings of Ref.~\cite{Raffaelli:2026bhk}, which recently compared these models against NPIPE data. Our $\Delta\chi^2$ values and Bayes factors are slightly different, which we trace back to the fact that, unlike that analysis, we do not include CMB lensing data.

As for the exotic energy-injection model, we obtain a bound on the decay rate $\Gamma$. By inspecting the 2D posterior distribution in the $(z_{\rm inj},\Gamma)$ plane, we see that the constraint becomes indeed strong for $0.8\lesssim\log_{10}z_{\rm inj}\lesssim3.15$, and that they relax around $z_{\rm inj}\sim1000$, and $z_{\rm inj}=10$. In particular, the best fit is located at $z_{\rm inj}\sim1150$ redshift. As discussed in the previous Section, an injection around that redshift produces a small oscillatory residual with the characteristic acoustic frequency. Since the positions and amplitudes of the acoustic peaks are very well measured, such a residual is only allowed if its amplitude is small, and therefore it does not significantly improve the fit to the data. This is reflected in the relatively small value of the decay rate at the best fit, which corresponds to only a small correction to the CMB spectra, as can be seen from the blue lines in Figs.~\ref{fig:bestfit-log} and \ref{fig:bestfit-sharp}.

Finally, let us discuss the constraints on EDE. As already pointed out in Ref.~\cite{Efstathiou:2023fbn}, NPIPE sets tighter constraints on EDE than previous versions of the Planck likelihoods. We obtain a tight marginalized bound on $f_{\rm EDE}< 0.0387$ at 95\% CL, and see that the posterior peaks around $\log_{10}z_{\rm EDE}\sim3.3$. While the toy-model considered here is different from the Axion-EDE model in~\cite{Efstathiou:2023fbn}, we also confirm that such a strong bounds limits the degeneracy of $f_{\rm EDE}$ with $H_0$ and, as such, its capability to ease the $H_0$ tension. We have tested that our conclusions are stable with respect to leaving the other EDE parameters introduced in Section~\ref{sec:late} free to vary in the analysis. As can be seen from Table~\ref{tab:bestfit_parameters}, the best fit from candidate has a very small value for $f_{\rm EDE}$ and the improvement to the fit is insignificant with respect to $\Lambda$CDM.

{\bf Joint analysis of primordial and post-inflationary features.} Finally, we consider the joint analysis of primordial and post-inflationary features. Figs.~\ref{fig:constraints-sharp},~\ref{fig:constraints-log}~\ref{fig:constraints-ede-sharp},~\ref{fig:constraints-ede-log} clearly show that the posterior distributions of the relevant parameters---the amplitude and frequency of the primordial features, the decay rate and redshift of the energy injection, and the amplitude and redshift of the EDE injection---are only minimally affected by the joint analysis and remain very close to those obtained in the corresponding single-model analyses.\footnote{The small visual differences between the posterior contours in Figs.~\ref{fig:constraints-sharp},~\ref{fig:constraints-log}~\ref{fig:constraints-ede-sharp},~\ref{fig:constraints-ede-log} are sensitive to the degree of smoothing. Stronger smoothing largely suppresses these differences, but also washes out the multimodal structure that we choose to preserve.}
The posterior distributions of the primordial-feature phases change slightly, but these parameters remain largely unconstrained in both the individual and joint analyses.
In particular, the addition of the second class of features does not open new degenerate directions in parameter space, significantly relax the constraints on the amplitudes, or shift the preferred regions in frequency or injection redshift. Interestingly, the primordial feature amplitude--frequency 2D posterior distributions retain the same multimodal structure regardless of whether the exotic energy injection or EDE is jointly analyzed or not. This provides a direct indication that the local high-likelihood regions associated with primordial features are not being absorbed or reproduced by the post-inflationary signal. Furthermore, we see that the parameters of the best fit candidates do not shift in the two scenarios.

We also note that, while in principle a small correlation can arise between a primordial feature with a frequency close to the acoustic one and an exotic energy injection around the epoch of recombination, this weak degeneracy does not lead to an appreciable correlation in the posterior distributions inferred from the data. 

Finally, let us note that the posterior distributions of the standard cosmological parameters are also stable when moving from the individual to the joint analyses.

The main result of this Section is therefore a confirmation of the conclusions of our PCA analysis: primordial features do not exhibit significant degeneracies with the classes of late-time signals considered here, be they exotic energy injections or an additional dark fluid in the form of EDE, and leave characteristic imprints in the CMB that cannot in general be reproduced by the post-inflationary signals. While these models are not significantly detected---see their Bayes factors in Tab.~\ref{tab:bestfit_parameters}, which penalize the additional parameter volume despite the improvement in the best fit---they do provide an improvement to the fit and may therefore still be lurking in the data. 
If such signals are detected by future experiments, our results indicate that their primordial or post-inflationary origin can likely be distinguished. 
Moreover, if the signal is primordial, different classes of primordial features can themselves be distinguished, as shown in Ref.~\cite{Braglia:2022ftm}.

\section{Conclusion}
\label{sec:conclusion}

In this work, we have investigated the prospects for distinguishing between primordial and post-inflationary origins of scale-dependent features in the CMB power spectrum. 
We focused on sharp and resonant primordial features and compared them to post-inflationary features arising from exotic energy injection and EDE-like models at various redshifts.

Our analysis reveals a clear qualitative distinction between these two broad classes of signals. 
Primordial features can imprint oscillatory residuals in the CMB power spectrum with essentially arbitrary frequency, determined by the momentum scale $k_0$ (for sharp features) or $\omega/H$ (for resonant features) characterizing the inflationary dynamics. 
In contrast, the post-inflationary signals studied here are limited to two main types of effects: an overall damping of the power spectrum through enhanced optical depth (although the exotic features have some additional mild $\ell$-dependence), or a phase shift corresponding to modified recombination timing or altered expansion history. 
Residuals in the CMB power spectrum due to the latter manifest as oscillations at the BAO frequency, providing a natural characteristic scale for post-inflationary signals.

We verified these findings through a principal component analysis, which demonstrates that the dominant directions in parameter space associated with primordial and post-inflationary features are nearly orthogonal. 
The PCA also confirms that primordial feature and EDE-like signals cannot be captured by a small number of templates, in the former case because the primordial oscillation frequency can vary freely, while exotic energy injection effects are well-described by just a few principal components. 
The only exception to this clean separation occurs when primordial features happen to have characteristic frequencies overlapping with the BAO scale.
However, in the case of exotic energy injection, the degeneracy is only partial and mild, as the non-trivial transfer function linking primordial to CMB features deforms the primordial oscillations in ways that remain distinguishable from pure post-inflationary effects.
For joint analyses with EDE-like models, there are no significant degeneracies with primordial features.

Our analysis with real CMB data from the Planck satellite provides a more quantitative proof of the intuitive results from the PCA analysis. A nested sampling analysis of the most recent and most stringent data release from the Planck collaboration---implemented in the CamSpec NPIPE likelihood---confirms that the posterior of both the primordial and the post-inflationary extra parameters do not change when analyzed jointly. 
These results are encouraging for future CMB observations. 
They suggest that if statistically significant residuals beyond $\Lambda$CDM are detected, it should generally be possible to attribute them to either primordial or post-inflationary physics based on the frequency content of the signal. 
The only ambiguous scenario, namely, oscillations at precisely the BAO frequency, would still favor a post-inflationary interpretation, as it would be disfavored from a Bayesian perspective, requiring fine-tuning of the primordial feature frequency and of the scale-dependent envelope of the oscillations.

We emphasize that our study represents a first step in systematically exploring this degeneracy problem. 
We have focused on specific, well-motivated examples: sharp and resonant features for the primordial sector, and energy injection and EDE-like models for post-inflationary physics. 
Many other exotic scenarios exist that could affect the CMB—including modifications to gravity, changes in recombination physics, or interactions between dark matter and dark radiation. 
However, the PCA and nested sampling methodologies employed here are general and can be straightforwardly extended to incorporate these additional classes of signals, provided one has access to a modified Boltzmann code (such as CLASS) that can compute their impact on the CMB power spectrum.
The qualitative distinction in how primordial and post-inflationary features affect the CMB, which we have learned from this study, can also be more broadly applicable.

Future high-sensitivity CMB experiments will be crucial not only for detecting or ruling out specific feature models, but also for discriminating between different primordial mechanisms \cite{Braglia:2022ftm,Petretti:2024mjy} and distinguishing them from post-inflationary alternatives. 
An important future direction is to extend this analysis beyond the CMB, since scale-dependent features may also have correlated imprints in large-scale structure. 
Current LSS constraints are competitive with CMB limits only for a limited range of feature frequencies, but are expected to surpass them soon for constraining primordial features~\cite{Huang:2012mr,Hazra:2012vs,Chen:2016vvw,Ballardini:2016hpi,Palma:2017wxu,LHuillier:2017lgm,Ballardini:2017qwq,Beutler:2019ojk,Ballardini:2019tuc,Debono:2020emh,Li:2021jvz,Ballardini:2022wzu,Chandra:2022utq,Chen:2026gxp}.
LSS will therefore provide additional handles for breaking degeneracies---thanks to the different transfer functions of LSS and CMB to these primordial and post-inflationary features---and test the consistency of primordial versus post-inflationary interpretations across multiple cosmological observables. 

\section*{Acknowledgments}

W.Q. was supported by the National Science Foundation Graduate Research Fellowship under Grant No. 2141064 and a grant from the Simons Foundation (Grant Number SFI-MPS-SFJ-00006250, W.Q.).  
MB has received funding from the European Union’s Horizon 2020
research and innovation programme under the Marie Skłodowska-Curie grant
agreement No 101205460. 
TRS' work is supported by the Simons Foundation (Grant Number 929255, T.R.S) and by the U.S. Department of Energy, Office of Science, Office of High Energy Physics of U.S. Department of Energy under grant contract Number DE-SC0012567.
This analysis made use of \texttt{Numpy} \cite{Harris:2020xlr}, \texttt{Scipy} \cite{2020NatMe..17..261V}, \texttt{Jupyter}~\cite{Kluyver2016JupyterN}, \texttt{tqdm}~\cite{daCosta-Luis2019}, and \texttt{Matplotlib} \cite{Hunter:2007ouj}. Some of the computations presented in this paper were performed on the FASRC Cannon cluster at Harvard University.

\appendix

\section{Analytic study of exotic energy injection}
\label{app:analytic}

In Section~\ref{sec:late}, we outlined the types of effects that exotic energy injection can have on the CMB power spectrum.
Here, we lay out the intuition behind these effects in more detail using a simplified analytic treatment.
To study the impact specifically on the acoustic oscillations, we can examine the evolution equation for photon density fluctuations $\delta_\gamma = \Delta \rho_\gamma / \bar{\rho}_\gamma$ while they are still coupled to baryons, which is given by~\cite{Dodelson:2020bqr,Baumann:2022mni}
\begin{equation}
    \delta_\gamma'' + \frac13 k^2 \delta_\gamma = - \frac43 k^2 \Psi + g(k, \eta) ,
\end{equation}
where $\eta$ denotes conformal time, primes denote derivatives with respect to $\eta$, $k$ stands for wavenumber, and $\Psi$ is the gravitational potential.
One might think that we would also have to include an additional forcing term to represent the effect of exotic injection, which we denote by $g(k, \eta)$; however, we will see momentarily that this is not the primary effect.

The solution to this equation is
\begin{equation}
    \delta_\gamma = - 4 \Psi + C \cos\left( \frac{k}{\sqrt{3}} \eta_\mathrm{rec} + \phi \right) + \frac{\sqrt{3}}{k} \int_0^{\eta_\mathrm{rec}} d\eta' \, g(k, \eta') \sin \left[ \frac{k}{\sqrt{3}} (\eta_\mathrm{rec} - \eta') \right]
    \label{eqn:modified_dgamma}
\end{equation}
where $C$ and $\phi$ are determined by the initial conditions (which are independent of energy injection), and we have evaluated the solution at the time of recombination $\eta_\mathrm{rec}$ when the acoustic oscillations are frozen in.
This $\eta_\mathrm{rec}$ can be defined either as when the optical depth $\tau$ is equal to one or where the visibility function (which is a function of $\tau$) peaks.
Hence, additional contributions to the optical depth from redshifts near $\eta_\mathrm{rec}$ may delay recombination to a new time, $\eta_\mathrm{rec} + \Delta \eta_\mathrm{rec}$.
If $\Delta \eta_\mathrm{rec}$ is small, then the change to the photon fluctuations is
\begin{equation}
    \Delta \delta_\gamma = - C \sin \left( \frac{k}{\sqrt{3}} \eta_\mathrm{rec} + \phi \right) \times \frac{k}{\sqrt{3}} \Delta \eta_\mathrm{rec} .
\end{equation}
This difference will appear as a phase shift with a mild scale dependence.

Since we are studying \textit{homogeneous} energy injection, the impact of the last term in Eqn.~\eqref{eqn:modified_dgamma} is unobservable in the CMB power spectrum.
To see this, consider $g(k,\eta)$ at a fixed value of $\eta$.
This forcing term must be the Fourier transform of a function that is constant across space, which means the $k$-dependence of this term is that of a delta-function, $\delta^{(3)} (\boldsymbol{k})$.
Therefore, this term only affects the zero-wavenumber mode for homogeneous energy injection (the mean temperature) and will not appear in the CMB anisotropies.

If we had also included in this derivation the collision terms due to Compton scattering, we would find the usual $e^{-\tau}$ dependence of the photon anisotropies.
Since exotic energy injection generally gives positive contributions to $x_e (z)$, this raises the value of $\tau$ and causes an overall damping of the CMB power spectrum.
The impact of exotic ionization is not completely degenerate with $\tau$, since the former usually carries some $\ell$-dependence based on the exact redshift history of the exotic injection.
The effect has been extensively discussed in the literature and used to constrain various models of exotic energy injection, see e.g. Refs.~\cite{Padmanabhan:2005es,Slatyer:2009yq}.

\section{PCA-based constraint on energy injection}
\label{app:PCA_injection}
\begin{figure*}
    \centering
    \includegraphics[width=\textwidth]{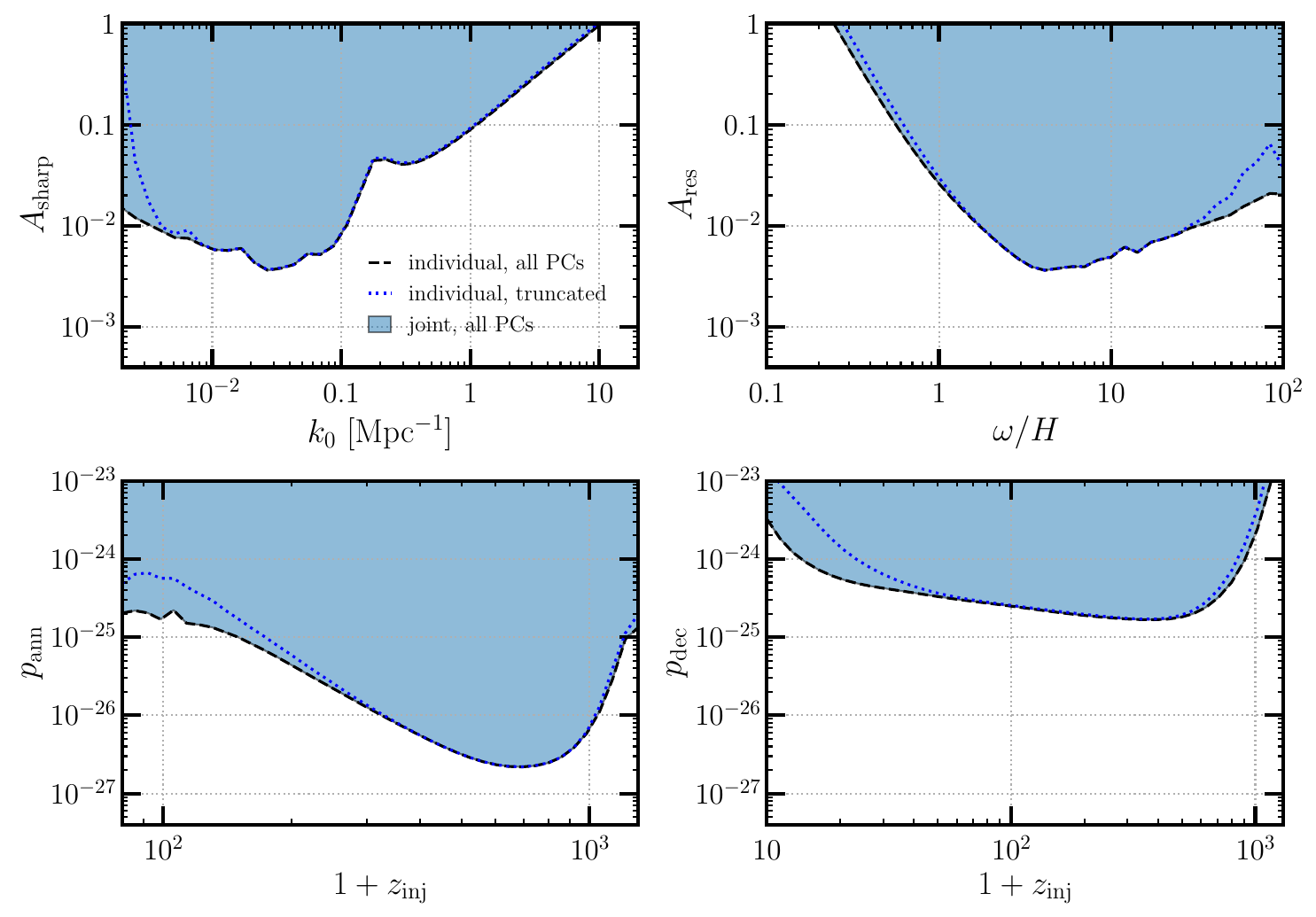}
    \caption{
    	Combined sensitivity estimates on inflationary features and post-inflation energy injection.
    	Going clockwise starting from top-left, the panels show the constraints on sharp inflationary features, resonant inflationary features, injections with a decay-like redshift dependence, and injections with an annihilation-like redshift dependence.
    	The injection results are broadly in agreement with past results from Ref.~\cite{Finkbeiner:2011dx} which assumed a WMAP-like instrument.
    }
    \label{fig:PCA-limits}
\end{figure*}

If primordial and post-inflationary features have orthogonal effects on the CMB, then the constraints on these models should not change significantly when including only one class of models vs.~both classes.
This is demonstrated in Fig.~\ref{fig:PCA-limits}, which shows the constraints on inflationary features and late-time energy injections from several different analyses, after marginalizing over cosmological parameters.
The black dashed line shows the result of setting the limits when including only one class of features and all PCs.
We also show the limit when only including PCs with eigenvalues greater than 1\% of the total constraining power in the blue dotted line; this curve well approximates the all-PC limit, especially where the constraint is strongest.
To improve the approximation across a broader range of basis models, one need only include more PCs in Eqn.~\eqref{eqn:PCA_limit}.
The blue contour shows the limits resulting from the joint feature constraints.
In principal, each panel should show two joint constraints, e.g. the the top left panel should show limits on sharp features from joint constraints with annihilation-like injections, and then a separate limit from joint constraints with decay-like injections.
However, we find in all cases that joint limits on the same classes of models are indistinguishable, so we show only one in each panel.

The results for annihilation/decay-like injections are broadly consistent with the results of Ref.~\cite{Finkbeiner:2011dx}---differences can be attributed to the fact that we are now assuming noise parameters for Planck, as opposed to WMAP.
For example, Ref.~\cite{Finkbeiner:2011dx} found that their annihilation constraints were largely determined by the level of exotic ionizations at $1+z \sim 600$ using WMAP 7 data---with Planck, we find the most constraining redshift to instead by $1+z \sim 700$.
Ref.~\cite{Finkbeiner:2011dx} also found that constraints on decaying dark matter are mostly controlled by ionizations at $1+z \sim 300$; however, we find that with the Planck noise parameters, the redshift to which the CMB fluctuations are most sensitive is shifted to $1+z \sim 400$.

\bibliographystyle{JHEP}
\bibliography{refs}
\end{document}